\documentclass[fleqn,usenatbib]{mnras}

\usepackage[T1]{fontenc}
\usepackage{ae,aecompl}

\usepackage{graphicx}	
\usepackage{amsmath}	
\usepackage{amssymb}	
\usepackage{gensymb}
\usepackage{txfonts}
\usepackage[dvipsnames]{xcolor}

\title[White dwarf pollution from resonances]{On the role of resonances in polluting white dwarfs by asteroids}

\author[Smallwood et al.]{
Jeremy L. Smallwood,$^{1}$\thanks{E-mail: Smallj2@unlv.nevada.edu} Rebecca G. Martin,$^{1}$ Mario Livio$^{1}$ and Dimitri Veras$^{2,3,4}$
\\
$^{1}$Department of Physics and Astronomy, University of Nevada, Las Vegas, 4505 South Maryland Parkway, Las Vegas, NV 89154, USA \\
$^{2}$Centre for Exoplanets and Habitability, University of Warwick, Coventry CV4 7AL, UK \\
$^{3}$Department of Physics, University of Warwick, Coventry CV4 7AL, UK \\
$^{4}$STFC Ernest Rutherford Fellow
}

\date{Accepted XXX. Received YYY; in original form ZZZ}

\pubyear{2018}

\begin{document}
\label{firstpage}
\pagerange{\pageref{firstpage}--\pageref{lastpage}}
\maketitle

\begin{abstract}
 Pollution of white dwarf atmospheres may be caused by asteroids that originate from the locations of secular and mean-motion resonances in planetary systems.  Asteroids in these locations  experience increased eccentricity, leading to tidal disruption by the white dwarf. We examine how the $\nu_6$ secular resonance shifts outwards into a previously stable region of the asteroid belt, as the star evolves to a white dwarf. 
Analytic secular models require a planet to be engulfed in order to shift the resonance. 
We show with numerical simulations that as a planet gets engulfed by the evolving star, the  secular resonance shifts and the rate of tidal disruption events increases with the engulfed planet's mass and its orbital separation.
We also investigate the behaviour of mean-motion resonances. The width of a mean-motion resonance increases as the star loses mass and becomes a white dwarf.  The $\nu_6$ secular resonance is  more efficient at driving tidal disruptions than  mean-motion resonances with Jupiter. By examining 230 observed exoplanetary systems whose central star will evolve into a white dwarf,  we find that along with an Earth mass planet at $1\,\rm au$, hot Jupiters at a semi--major axis $a\gtrsim 0.05\,\rm au$  and super--Earths of mass $10\,\rm M_\oplus$ at  $a\gtrsim 0.3\,\rm au$  represent planet types whose engulfment shifts resonances enough to cause pollution of the white dwarfs to a degree in agreement with observations.
\end{abstract}




\section{Introduction}

The majority of the stars in the Milky Way (more than $97$ per cent) will evolve to become a white dwarf \citep{Althaus2010}. Spectroscopic observations reveal metallic absorption lines in the atmospheres of white dwarfs \cite[e.g.,][]{Zuckerman2007,Klein2010,Vennes2010,Zuckerman2010,Farihi2012,Melis2017,xu2018b,Hollands2018,Harrison2018,Doyle2019,Swan2019,Bonsor2020,Doyle2020,Harrison2021,Kaiser2021,Klein2021}. Around $20$ to $50$ per cent of all white dwarfs show traces of metal-polluted atmospheres \citep{Zuckerman2003,Zuckerman2010,Koester2014}. The source of this pollution is thought to be associated with the white dwarf disrupting and accreting asteroids or small bodies from their primordial planetary systems \citep{Debes2002,Jura2003}. 

The metals eventually sink and diffuse within the atmosphere due to the WD's intense surface gravity \citep{Fontaine1979,Vauclair1979,Koester2006,Koester2009,Bauer2019,Blouin2020,Cunningham2021}. The stratification timescale for metals is of the order of days to a few Myr, depending on the composition of the white dwarf's atmosphere \citep{Koester2006}. Since the diffusion timescale of metals is orders of magnitude shorter than the WD cooling ages \citep{Paquette1986,Koester2009}, polluted white dwarfs must be continuously accreting metal-rich material. For a recent review on the dynamics of white dwarf pollution, see \cite{Veras2016}.

The prevailing scenario for the provision of long term  accretion of metal-rich material is that planetary debris are excited onto star-grazing orbits and ultimately become disintegrated by tidal forces, forming a debris disc around the white dwarf \citep{Gansicke2006,Kilic2006,vonHippel2007,Farihi2009,Jura2009,Farihi2010b,Melis2010,Brown2017,Bonsor2017,xu2018a,Debes2019,Wilson2019}. The planetary debris that have been hypothesized as a source include asteroids \citep{Jura2003,Jura2006,Jura2009,Debes2012,Veras2013b,Wyatt2014,Frewen2014,smallwood2018b, Mustill2018,VerasHiguchi2019,Valeri2019,Martin2020}, comets \citep{VerasShannon2014,Stone2015,Caiazzo2017}, and moons \citep{Payne2016,Payne2017}. There is also evidence for pollution of the atmospheres of white dwarfs in close-in binaries which contain a circumbinary debris disc \citep{Farihi2017}. Material originating from the interstellar medium has been ruled out as a source of pollution \citep{Aannestad1993,Jura2006,Kilic2007,Farihi2010,Barstow2014}.

The mechanisms that could potentially drive white dwarf pollution include mean-motion resonances \citep{Debes2012,Voyatzis2013}, secular resonances \citep{smallwood2018b} and planet-planet scattering \citep{Payne2016,Payne2017}. Moreover, the Kozai-Lidov instability may also provide pollution within binary systems \citep{Hamers2016,Petrovich2017,Stephan2017,Stephan2018}. Recently, \cite{VerasXu2018} computed the critical separation of binaries required for the atmosphere of a white dwarf to become polluted by Roche lobe overflow or by stellar winds. The critical separation is only a few astronomical units \cite[roughly agreeing with Fig.3 in][]{Debes2006}, which implies that other mechanisms are needed for wide-binary systems \citep{Kratter2012,Bonsor2015}.

Direct observational evidence for the proposed scenarios is still scarce but more are on the horizon \cite[e.g.,][]{Guidry2020}. Currently, there are only two known white dwarfs to exhibit periodic transits of planetary debris, WD 1145+017 \citep{Vanderburg2015,Xu2016} and ZTF J0139+5245 \citep{Vanderbosch2020}. WD 1145+017 also exhibits absorption lines from an eccentric circumstellar gas \citep{Xu2016,Redfield2017} and transits of debris fragments \citep{Gansicke2016,Rappaport2016,Gary2017,xu2018a}. \cite{Veras2017}, \cite{Duvvuri2020}, and \cite{OConnor2020} constrained the interior structure of the asteroid being disintegrated. The planetary debris in ZTF J0139+5245 has a period of $110\, \rm days$ and if the debris passes through the tidal disruption radius of the white dwarf, then the material has an estimated eccentricity of $>0.97$ \citep{Vanderbosch2020}. Even if not, disruption of the progenitor likely occurred within a few Roche radii of the white dwarf \citep{Veras2020a}, still suggesting a highly eccentric orbit. It is estimated that $1\%$ to $4.5\%$ of white dwarfs display an infrared excess from dust \citep{Becklin2005,Kilic2006,Jura2007,RebassaMansergas2019,Wilson2019}. The number of gaseous debris discs around polluted white dwarfs have increased to about $21$ \citep{Dennihy2018,Manser2020,GentileFusillo2020,Melis2020,Dennihy2020}.
Furthermore, a ferrous core fragment has been discovered orbiting SDSS J1228+1040 \citep{Manser2019,Bromley2019,Grishin2019,Veras2019,OConnor2020}.

 The sequence of events leading to pollution by asteroids is thought to be as follows. As a star's outer envelop expands, close-in planets are engulfed \citep{Siess1999,Villaver2007,Villaver2009,Mustill2012,Adams2013,Villaver2014,Ronco2020}, causing dynamical changes to the system \citep{Duncan1998}. The semimajor axis of the surviving objects expands adiabatically, 
 due to the mass loss from the star \citep{Reimers1977,McDonald2015,Rosenfield2014,Rosenfield2016}. The mass loss rate can be as high as $10^{-4}\, \rm M_{\odot}yr^{-1}$ \citep{Veras2011}. \cite{smallwood2018b} used secular theory and $N$-body simulations to show that in the case of the Solar system, as the Earth is engulfed during stellar evolution \citep{Schroder2008}, the $\nu_6$ secular resonance shifts outwards relative to the asteroid belt, into a previously stable region of the asteroid belt. Secular resonances occur when the free apsidal precession frequency of two objects are equal. As the resonance location shifts outwards, it excites the eccentricities of  these formerly stable asteroids thus causing the debris to be perturbed onto star-grazing orbits and eventually to become tidally disrupted by the white dwarf. The disrupted material then forms an accretion disc, which subsequently pollutes the white dwarf atmosphere.  Secular resonances can supply a steady pollution of debris to the tidal disruption radius of the white dwarf over Myr timescales, which is roughly equivalent to the lifetime of the debris discs \citep{Girven2012,Veras2020b}.

In the present paper, we extend the work of \cite{smallwood2018b} by examining the effects of the mass and orbital semi--major axis of the engulfed planet on the secular resonance shift. We assume that the planets and the asteroid belt are sufficiently far from the white dwarf to survive through the red-giant branch and the asymptotic giant branch phases. In Section~\ref{sec:sec_theory} we summarize the secular perturbation theory and examine the effect of the mass and orbital semi--major axis of the engulfed planet on the secular resonance shift.
In Section~\ref{sec:Nbody} we describe simulations in which we consider two initially narrow belts of asteroids, the first centered on the $\nu_6$ resonance and the second centered on the 2:1 mean motion resonance with Jupiter. We consider the effects of engulfed inner planets with an Earth mass and three Earth masses. In Section~\ref{sec:exoplanet}, we consider white dwarf pollution in other observed exoplanetary systems assuming that they have two giant outer planets (similar to Jupiter and Saturn in the solar system) and an asteroid belt. Although it is difficult to detect giant planets at large orbital radii, this configuration may be common \citep[e.g.][]{Martin2015}. The snow line radius in a protoplanetary disc is the orbital radius outside of which ice forms \citep{Lecar2006,Martin2012snowline,Martin2013}. Giant planets are expected to form outside the snow line radius in a protoplanetary disc due to the increased solid mass density there \citep{Pollack1996}.  The occurrence rate of Jupiter analogous around solar-type stars is estimated to be $6\%$ \citep{Wittenmyer2016}. However, the occurrence rate of giant planets rises for the stellar masses typical of WD progenitors \citep{Reffert2015}, which is around $2\, \rm M_{\odot}$ \citep{Koester2014,Tremblay2016}.  Asteroid belts are a result of the increased eccentricity of planetesimals inside of the orbit of  a giant planet \citep[e.g.][]{Morales2011} and thus asteroid belts likely coincide with the location of the snow line radius \citep{Martin2013asteroids}. Finally, we draw our conclusions in Section~\ref{sec:conc}.

\begin{figure}
\includegraphics[width=\columnwidth]{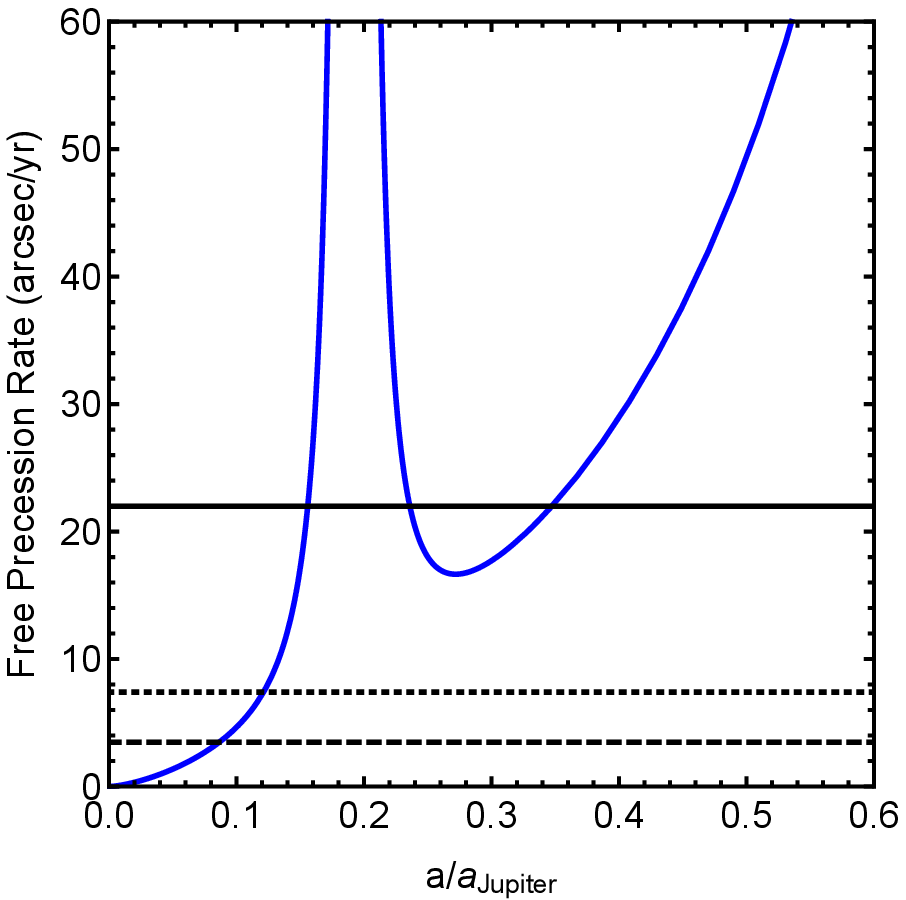}
\includegraphics[width=\columnwidth]{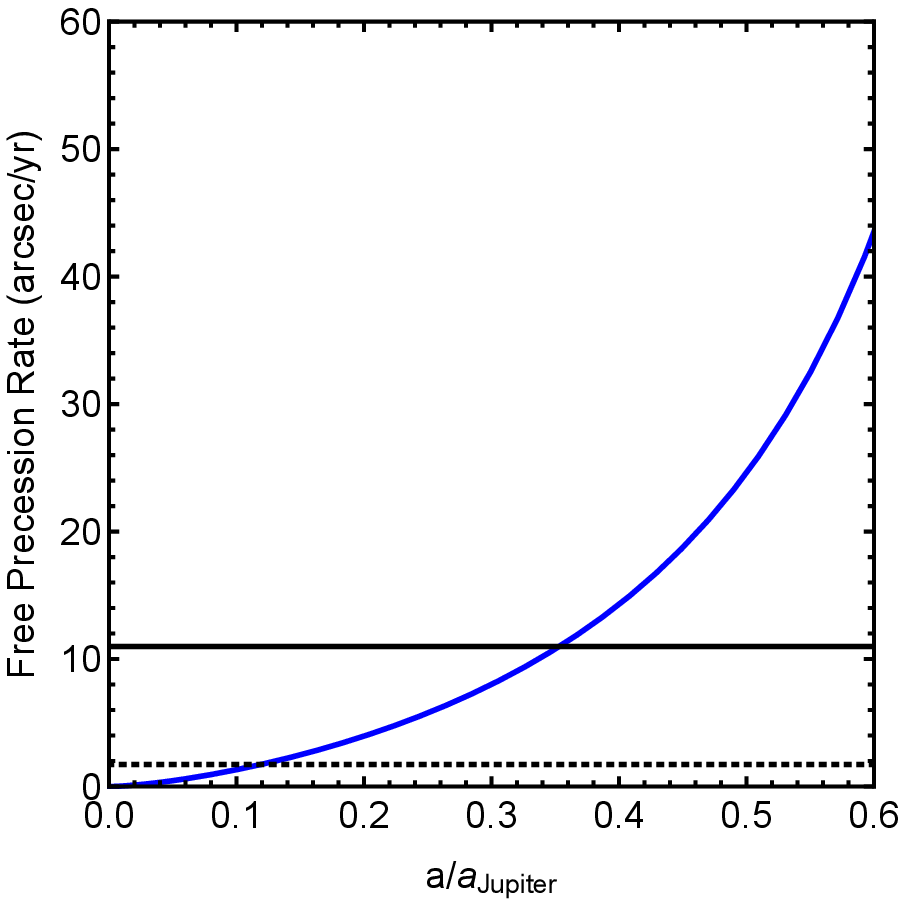}
\caption{The free apsidal precession rate of a test particle (blue) as a function of normalized semi-major axis from the analytic theory. The horizontal lines denote the eigenfrequencies of Earth (dashed), Jupiter (dotted), and Saturn (solid). Top panel: main-sequence stage with Earth, Jupiter, and Saturn. Bottom panel: white dwarf stage with only Jupiter and Saturn. The semi--major axis of Jupiter is $a_{\rm Jupiter}=5.2\,\rm au$ in the upper panel and $a_{\rm Jupiter}=10.4\,\rm au$ in the lower panel.}
\label{prec}
\end{figure}


\section{Secular Theory}
\label{sec:sec_theory} 
In this section, we generally follow the secular theory described by \cite{smallwood2018b}.  The Laplace-Lagrange equations are used to calculate the eigenfrequencies of $N$ planets and  the free apsidal precession rate of a test particle in the potential of the planetary system. The mean apsidal precession frequency, $g_0$, corresponds to the diagonal term of the Laplace-Lagrange matrix including the test particle \citep{Milani1990,morbidelli1991}. The analytical model we use is expanded to second order in eccentricity and inclination, which indicates that the secular perturbations are to second order in the orbital perturbation. The circumstellar radii at which the free apsidal precession frequencies of test particles are equal to any proper mode of $N$ planets determines the radial location of secular resonances. In the solar system, the $\nu_6$ secular resonance shapes the inner edge of the asteroid belt at about $2\, \rm au$. At this location, the free apsidal precession frequency of the asteroids is equal to the proper mode dominated by Saturn.

\begin{figure}
    \centering
    \includegraphics[width=0.95\columnwidth]{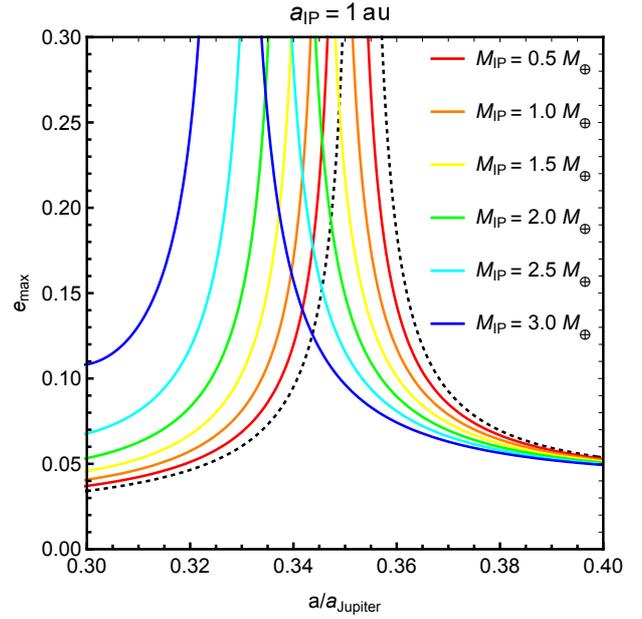}
     \caption{The maximum forced eccentricity as a function of the semi-major axis of a test particle near the $\nu_6$ secular resonance during the main-sequence stage (solid lines) versus the white dwarf stage (dotted-black line). The eccentricity during the main-sequence is calculated with an inner planet, Jupiter, and Saturn. As the star evolves towards the white dwarf stage, the inner planet is engulfed.  The colored lines show a different initial mass of the inner planet, $M_{\rm IP}$, as indicated by the legend. The inner planets' semi-major axis is fixed to $a_{\rm IP}=1\, \rm au$. The region of high maximum eccentricity in the coloured lines that does not overlap with the region of high maximum eccentricity in the dotted lines represents the region where previously stable asteroids that undergo increased eccentricity growth due to secular resonant perturbations.  The analytic theory is not accurate for such high values of the eccentricities ($e\gtrsim 0.2$), but we show it as an illustration of the effect.}
    \label{width}
\end{figure}


We consider the secular resonance for a planetary system that initially consists of an inner planet, Jupiter, and Saturn. During the main-sequence stage,  Jupiter and Saturn are at their current orbital locations with their current masses. In the post main sequence stage we take as the mass of the white dwarf into which the Sun has evolved to be half a solar mass. This is consistent, within the uncertainties,  with the determined initial-final mass relation \citep{Cummings2018}. 
The inner planet has been engulfed leaving only Jupiter and Saturn that orbit at twice their current orbital separation. It should be noted that during the Sun's giant branch evolution, Jupiter and Saturn will not undergo any instabilities with Uranus and Neptune \cite[e.g.,][]{Veras2016b}.

We first consider a system in which the inner planet is an Earth mass planet orbiting at a semi-major axis of $a_{\rm IP}=1\,\rm au$. The top panel of Fig.~\ref{prec} shows the eigenfrequencies of the three planets during the main-sequence stage (given by the horizontal lines and starting from top to bottom are Saturn, Jupiter, and the Earth) and the free precession rate of the particle is given by the blue curve. The asymptotic feature at about $0.2 a/a_{\rm Jupiter}$ is the location of the Earth. The location of the $\nu_6$ resonance in the main-sequence phase is at the outermost crossing of the free precession rate and the eigenfrequency of Saturn. This is at an orbital radius of about $0.347 a_{\rm Jupiter}= 1.806\, \rm au$. The discrepancy between the secular theory and the observed position, $\sim 2\, \rm au$, stems from the theory being of second-order only, thereby neglecting higher-order terms within the disturbing function. The bottom panel of Fig.~\ref{prec} is identical but for the white dwarf stage, where the Earth had been engulfed.  The $\nu_6$ resonance occurs where the free precession rate crosses the eigenfrequency of Saturn at $a=0.353a_{\rm Jupiter}=3.675 \,\rm AU$, where $a_{\rm Jupiter}$ is now the adiabatically expanded semi-major axis of Juipter.  Thus,  the $\nu_6$ resonance is shifted outward by $0.006\,a_{\rm Jupiter}$. This means that in the adiabatically expanded asteroid belt, the resonance shifts by $0.06\, \rm au$ relative to the belt. These results are independent of the size distribution of the asteroid belt since it is argued that  the individual asteroids should break up into smaller pieces during the asymptotic giant branch evolution \cite[e.g.,][]{VerasJacobson2014,VerasScheeres2020}.


 Figure~\ref{width} shows the corresponding maximum forced eccentricity  of a test particle as a function of semi-major axis \citep[see, e.g., equation (16) in][]{smallwood2018b}. The maximum eccentricity for the main-sequence stage  and the white dwarf stage are given by the solid-coloured lines and the black-dotted line, respectively. We consider the effect of the mass of the inner planet, $M_{\rm IP}$, on the resonance shift. The various coloured lines show the maximum eccentricity for test particles when varying the mass of the engulfed planet for a fixed semi-major axis of the engulfed planet of $a_{\rm IP}=1\,\rm au$.  The region of high maximum eccentricity in the coloured lines that does not overlap with the region of high eccentricity in the dotted-line indicates where previously stable asteroids undergo higher eccentricity growth by secular resonant perturbations due to the engulfment of the Earth-like planet. The more massive the inner planet, the wider the region of high eccentricity growth and the smaller the radius of the orbital location of the resonance in the main-sequence stage. The maximum eccentricity in the post-main sequence stage is the same independently of the mass of the engulfed planet. Hence, the dotted line is identical for each case. The more massive the engulfed planet, the more significant the secular resonance shift.



\begin{figure}
    \centering
    \includegraphics[width=0.95\columnwidth]{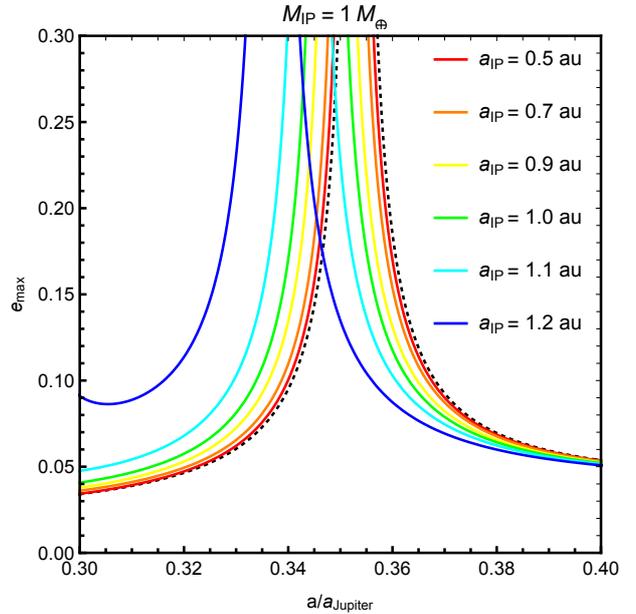}
    \caption{Same as Fig.~\ref{width} but with a fixed inner planet mass, $M_{\rm IP}=1\, \rm M_{\oplus}$, at different initial values of semi-major axis of the inner planet, as indicated by the legend. }
    \label{width_a}
\end{figure}

We also explored how the semi-major axis of the inner planet affects the resonance shift. Fig.~\ref{width_a} shows the maximum eccentricity for test particles when varying the semi-major axis of the engulfed planet for a fixed mass of the engulfed planet of $1\,\rm M_{\oplus}$. The larger the semi-major axis of the inner planet, the broader the region of high eccentricity growth and the smaller the radius of the orbital location of the resonance in the main--sequence stage. As the engulfed planet's orbital separation increases so does the resonance shift distance.

 The forced eccentricity and hence the location and width of the secular resonance is unchanged if the planet and star masses are all changed by the same factor. This means that, unlike mean-motion resonances, secular resonances don't broaden with stellar mass loss. However, systems of lower-mass outer planets with the same semi-major axis ratio should also be efficient at delivering planetary material through secular resonances. The time-scale for eccentricity excitation would be longer, but this may allow metal delivery to the white dwarf over longer time-scales without the belt being depleted \cite[e.g.,][]{Mustill2018}. The behaviour of the secular resonances beyond the linear regime may be different, reducing the delivery efficiency, but this could be tested with $N$--body simulations, which we show in the next section.

In the next section, we use $N$--body simulations to examine how the engulfed planet's mass relates to the resonance shift and we compare the effectiveness of the secular resonance to mean--motion resonances at driving tidal disruption events. 


\begin{table}\centering
 \caption{Simulation models denoting the evolutionary stage, main-sequence (ms) or white dwarf (wd), the mass of the inner planet ($M_{\rm IP}$), and whether the simulation modelled the $\nu_6$ secular resonance or the 2:1 mean-motion resonance. The models labeled with "no planet", are simulated without a inner planet.}
 \label{tab:Pconstrained}
 \begin{tabular}{lccc}
  \hline
  Model & ms/wd & $M_{\rm IP}/M_{\oplus}$  & $\nu_6$/2:1\\
    \hline
    \hline
     v6\_no\_planet\_ms  & ms & --  & $\nu_6$    \\
    \hline
     v6\_no\_planet\_wd   & wd & --  & $\nu_6$    \\
    \hline
     v6\_1ME\_ms  & ms & $1$ & $\nu_6$ \\
    \hline
     v6\_1ME\_wd  & wd & -- & $\nu_6$\\
    \hline
     v6\_3ME\_ms  & ms & $3$  & $\nu_6$   \\
    \hline
     v6\_3ME\_wd  & wd & --  & $\nu_6$    \\
    \hline
     21\_no\_planet\_ms  & ms & --  & $\nu_6$    \\
    \hline
     21\_no\_planet\_wd   & wd & --  & $\nu_6$    \\
    \hline
     21\_1ME\_ms & ms & $1$   & 2:1 \\
    \hline
     21\_1ME\_wd  & wd & --   & 2:1 \\
    \hline
     21\_3ME\_ms  & ms & $3$ & 2:1   \\
    \hline
     21\_3ME\_wd  & wd & -- & 2:1   \\
    \hline
 \end{tabular}
 \label{table::setup}
\end{table}



\section{$N$--body Simulations}
\label{sec:Nbody}
We use the symplectic integrator in the orbital dynamics package within {\sc mercury} \citep{Chambers1999} to simulate a planetary system with an asteroid belt during the main-sequence stage and then during the white dwarf stage (skipping the giant branch phases).  This is  a pure N-body setup, meaning we neglect any gas drag or radiation forces during the RGB/AGB phases. During the main-sequence phase, we assume the Sun, the inner planet (representing Earth), Jupiter, and Saturn to have the present-day orbital parameters, respectively. For the white dwarf models, the mass of the central object is halved, and the inner planet is removed from the simulation under the assumption that it is engulfed during stellar evolution.  All surviving bodies that have orbital radii well below a few hundred au of the star expand adiabatically \citep{Veras2013a}. This adiabatic expansion is relevant when the timescale for mass loss is much longer than the orbital periods of the surviving objects.

We explore the dynamics of the secular resonance shift by planetary engulfment for two masses of the inner  planet, $M=1\, \rm M_{\oplus}$ and $M=3\, \rm M_{\oplus}$. Based on the analytical approximation presented in Fig.~\ref{width}, we expect that the $3\, \rm M_{\oplus}$ case should produce more tidal disruption events compared to the $1\, \rm M_{\oplus}$ case.   \cite{Debes2012} found that as the mass of the central star decreases, the 2:1 mean motion resonance width increases and causes previously stable asteroids to become perturbed onto star-grazing orbits. Therefore, within each model, we also test the efficiency of  the frequency of tidal disruption  events produced from secular resonances versus mean-motion resonances. Table~\ref{table::setup} summarizes the simulation models. Each model is simulated for $100\, \rm Myr$. For the main-sequence models, this time is sufficient for the asteroid belt to approximate a steady state rate of asteroid clearing. The final conditions for the main-sequence models are then used as the initial conditions for the white dwarf models. We also include simulations for the $\nu_6$ and 2:1 resonances that do not initially include an inner planet (only Jupiter + Saturn).

Along with the planets, we consider a fiducial belt of test particles. The orbital elements for each particle are chosen as follows: the semi-major axis ($a$) was sampled uniformly in a range based on the type of resonance. For the $\nu_6$ resonance simulations, $1.7{\,\rm au} < a < 2.5{\,\rm au}$, and for the 2:1 resonance simulations, $3.15{\,\rm au} < a < 3.4{\,\rm au}$. The belts within each  model have the same density of particles per unit distance from the star. The initial number of particles is $\sim 10,000$ and $\sim 7500$ for the $\nu_6$ and 2:1 models, respectively. The inclination angle ($i$) is randomly distributed in the range $0-10\degree$, and the eccentricity ($e$) is randomly allocated from the range $0.0-0.1$. The remaining orbital elements, the longitude of the ascending node ($\Omega$), the argument of perihelion ($\omega$), and the mean anomaly ($M_{\rm a}$), were all randomly allocated in the range $0-360\degree$. The asteroids in our simulations are considered to be point particles that do not interact gravitationally with one another. We may neglect this interaction because the time-scale for an asteroid-asteroid collisional interaction is much longer than the time-scale for the action of perturbations by resonance effects. The time-scale for mean-motion resonant effects is of the order of $\sim 1\, \rm Myr$ \citep{ItoTanikawa1999}, whereas some of the largest asteroids have collisional time-scales that are of the order of the age of the Solar system \citep{Dohnanyi1969}.

\begin{figure}
\includegraphics[width=\columnwidth]{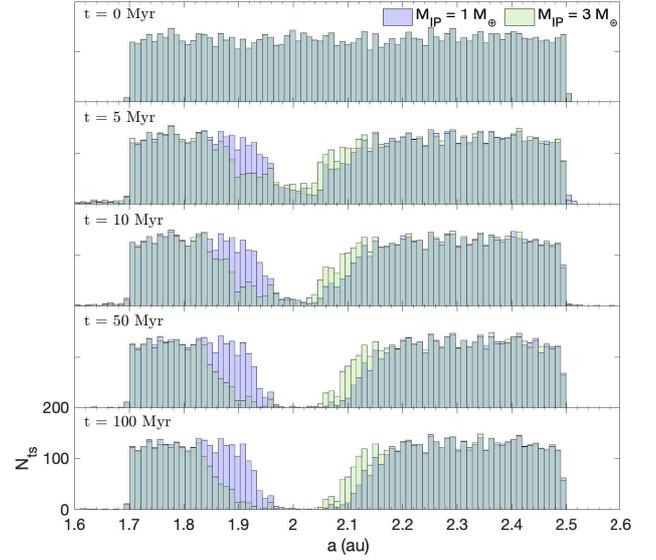}
\caption{ The number distribution of surviving test particles ($N_{\rm ts}$) as a function of semi-major axis at specific times for simulations of the $\nu_6$ secular resonance.  Blue corresponds to the model having a $1\, \rm M_{\oplus}$ inner planet (model v6\_1ME\_ms), while the green corresponds to the model with a $3\, \rm M_{\oplus}$  inner planet (model v6\_3ME\_ms). The gray regions represent the overlap between the two models. } 
\label{fig::ms_v6}
\end{figure}

\begin{figure}
\includegraphics[width=\columnwidth]{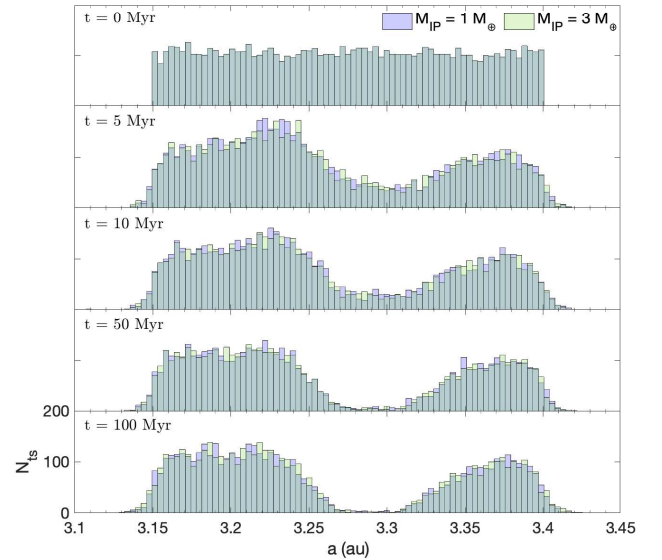}
\caption{Same as Fig.~\ref{fig::ms_v6} but for simulations around the 2:1 mean-motion resonance  (models 21\_1ME\_ms and 21\_3ME\_ms).  }
\label{fig::ms_21}
\end{figure}

\begin{figure*}
\includegraphics[width=1\columnwidth]{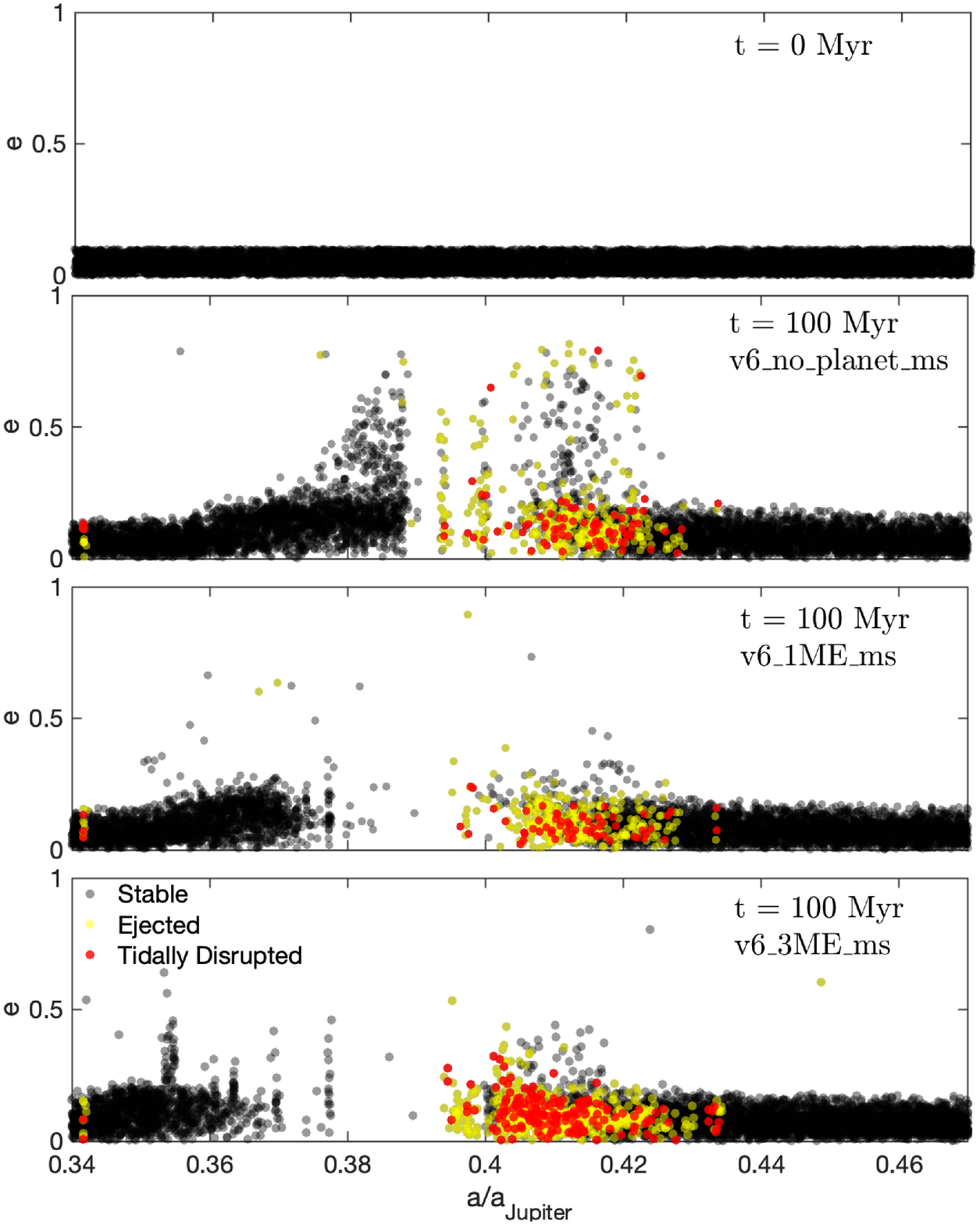}
\includegraphics[width=1\columnwidth]{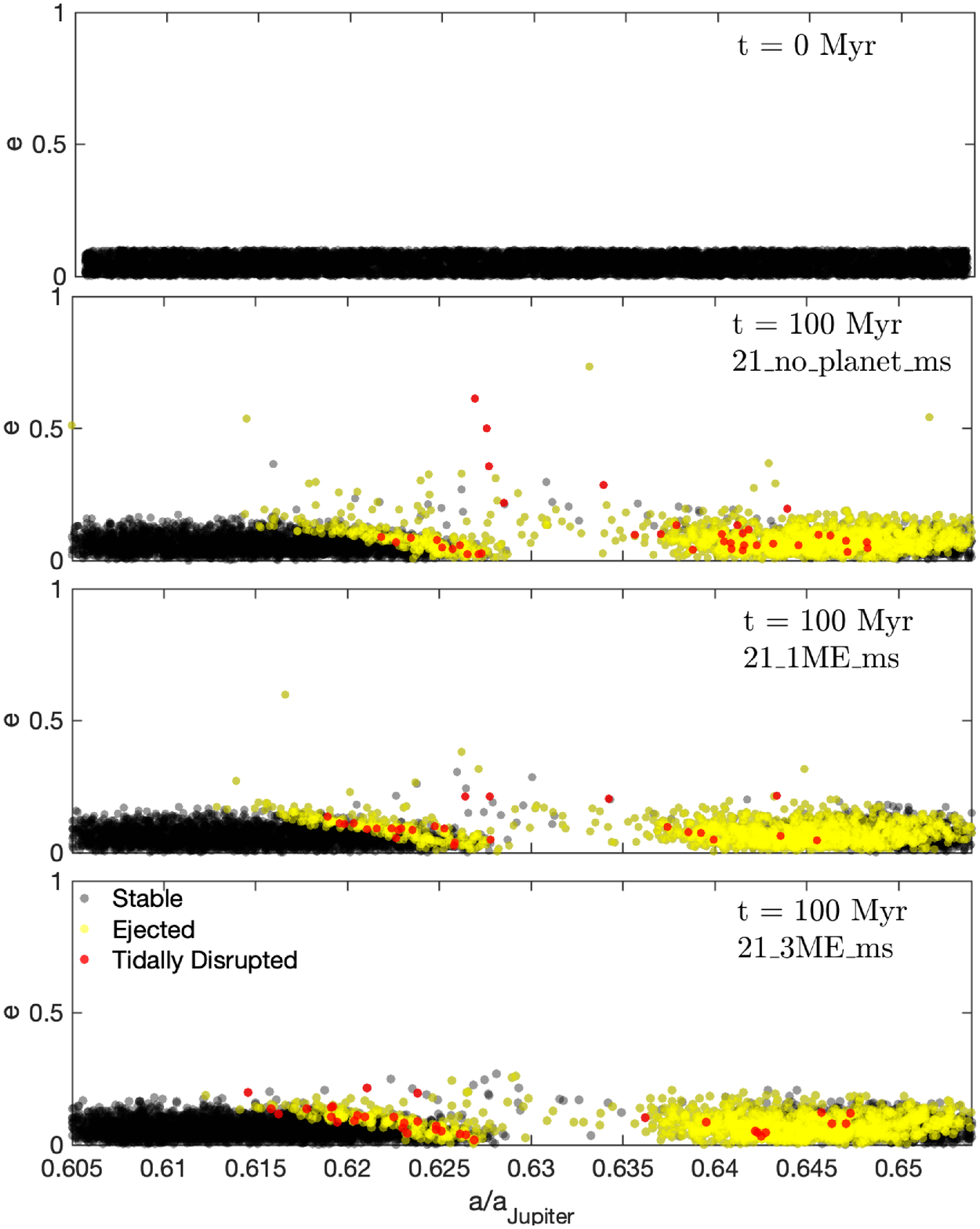}
\caption{The eccentricity versus semi--major axis of the asteroids at the end of the main-sequence stage. The left column shows the $\nu_6$ secular resonance simulations and the right column shows results of the 2:1 mean-motion resonance simulations.  The first row denotes the initial distribution of the belt. The second row shows the final distributions with no inner planet. The third row denotes the final distributions for belts with a $1\, \rm M_{\oplus}$ inner planet, while the fourth row denotes the final distributions for belts with a $3\, \rm M_{\oplus}$ inner planet. The colors denote the outcomes of the particles after the white dwarf stage. The outcomes include ejection (yellow), tidal disruption (red), or remaining stable (black). }
\label{fig::e_dist}
\end{figure*}

The possible outcomes for test particles near secular and mean-motion resonances include ejections, collisions with a larger body, or remaining within the simulation.  A particle is considered ejected if its semi-major axis exceeds $100\, \rm au$ and it is counted as a tidal disruption event if the particle passes within the white dwarf's tidal disruption radius. The tidal disruption radius for a $0.5\, \rm M_{\odot}$ white dwarf is $1.22\, \rm R_{\odot}$  with an asteroid density of $3\, \rm g/cm^{3}$ \cite[from equation~(17) in][]{smallwood2018b}.

\begin{figure*}\centering
\includegraphics[width=\columnwidth]{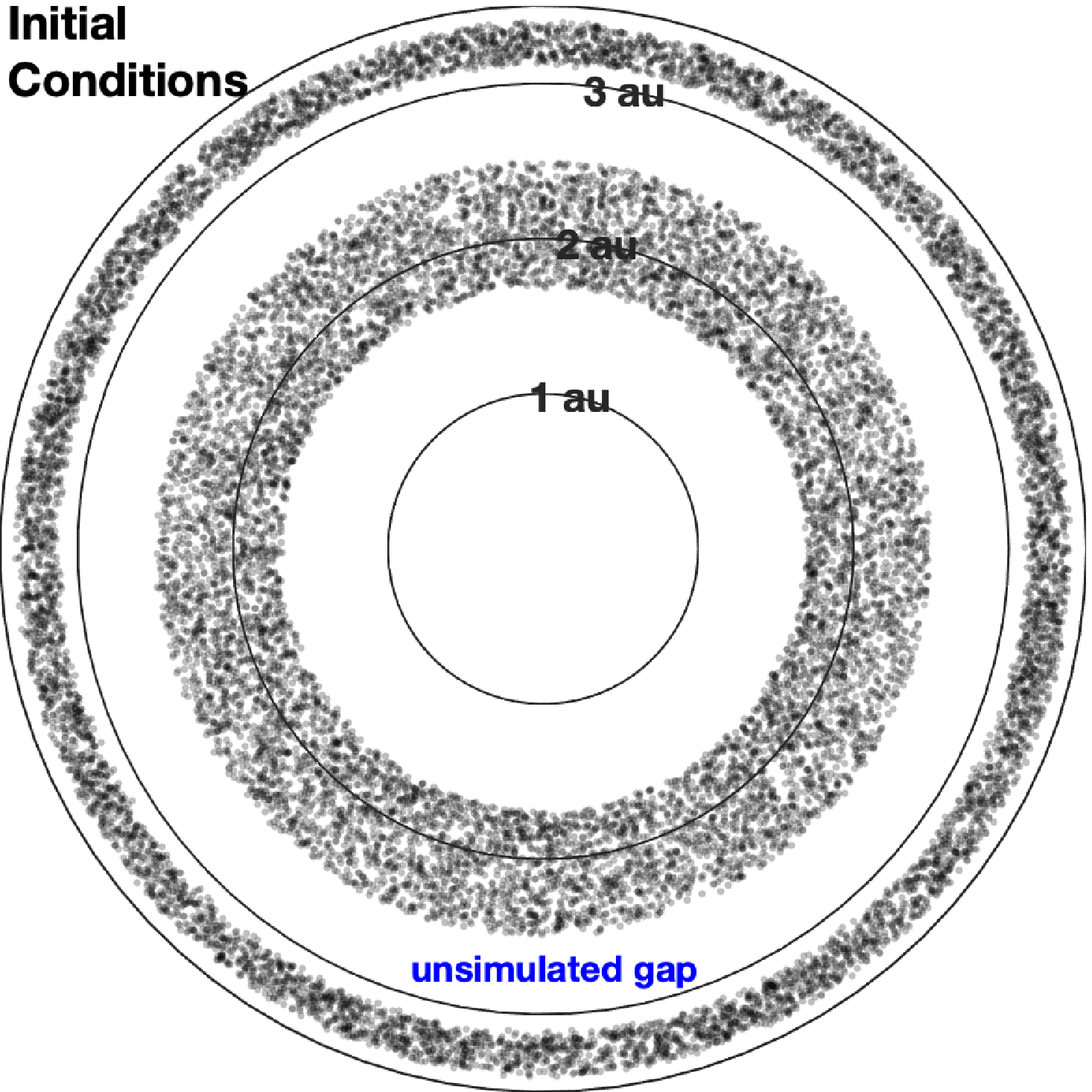}
\includegraphics[width=\columnwidth]{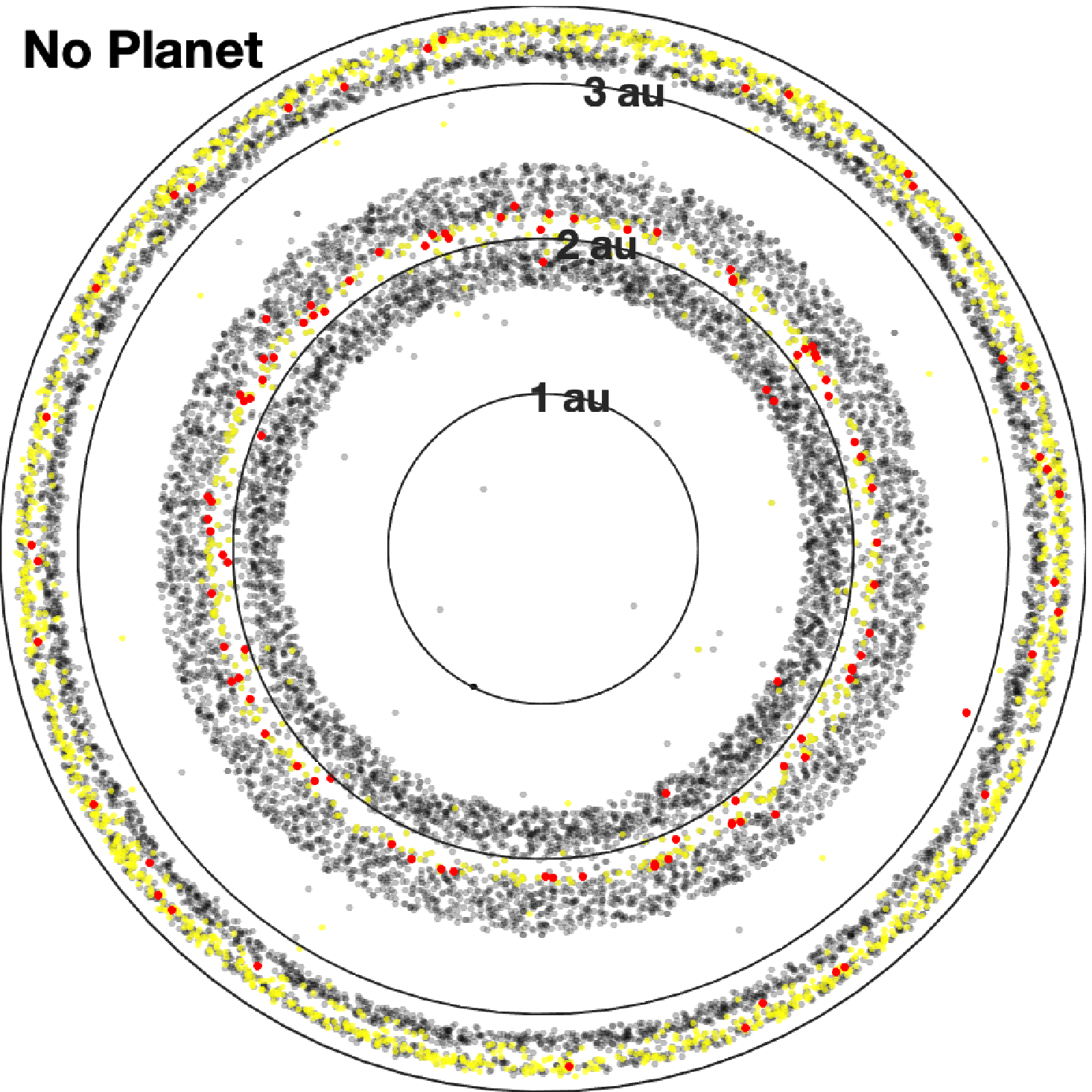}
\includegraphics[width=\columnwidth]{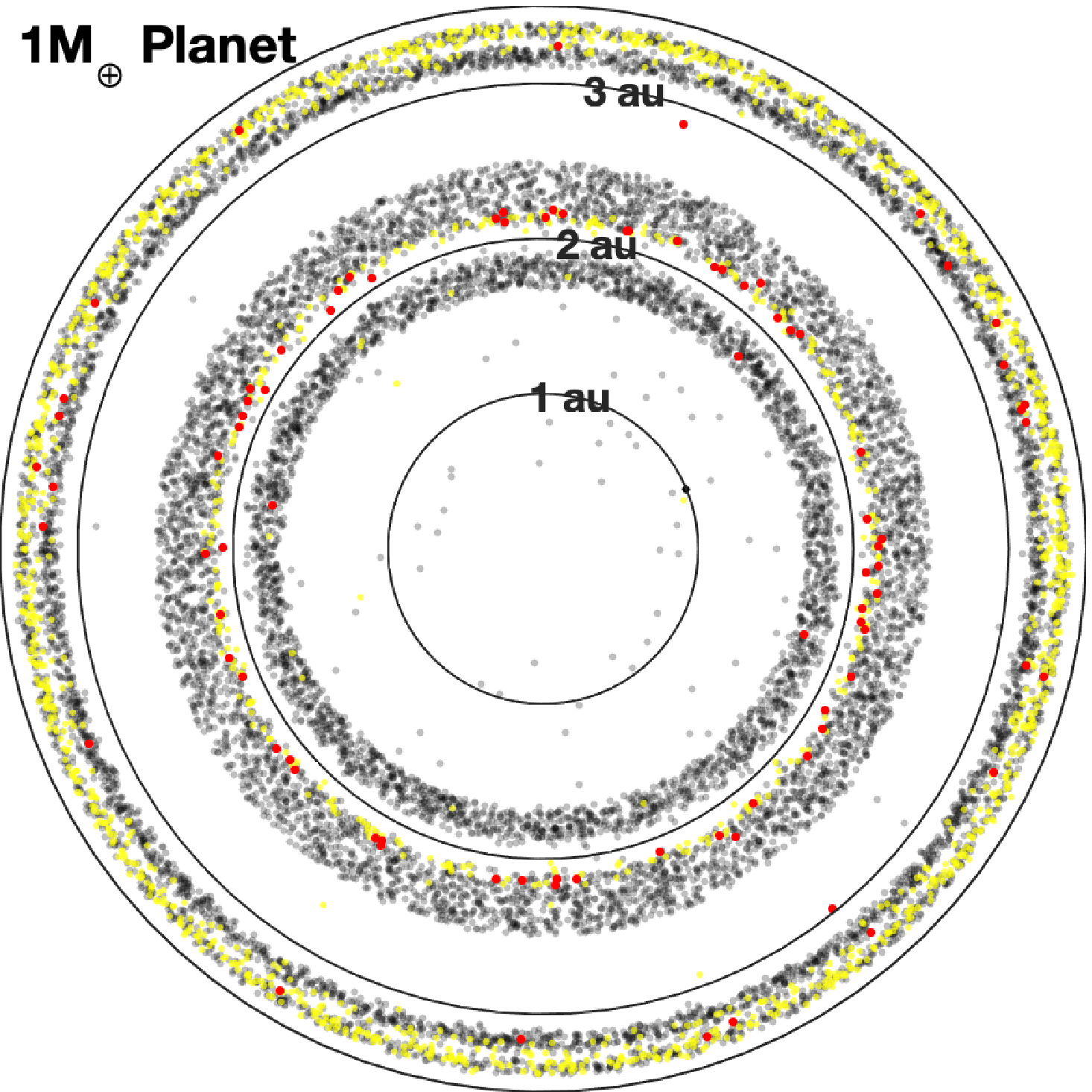}
\includegraphics[width=\columnwidth]{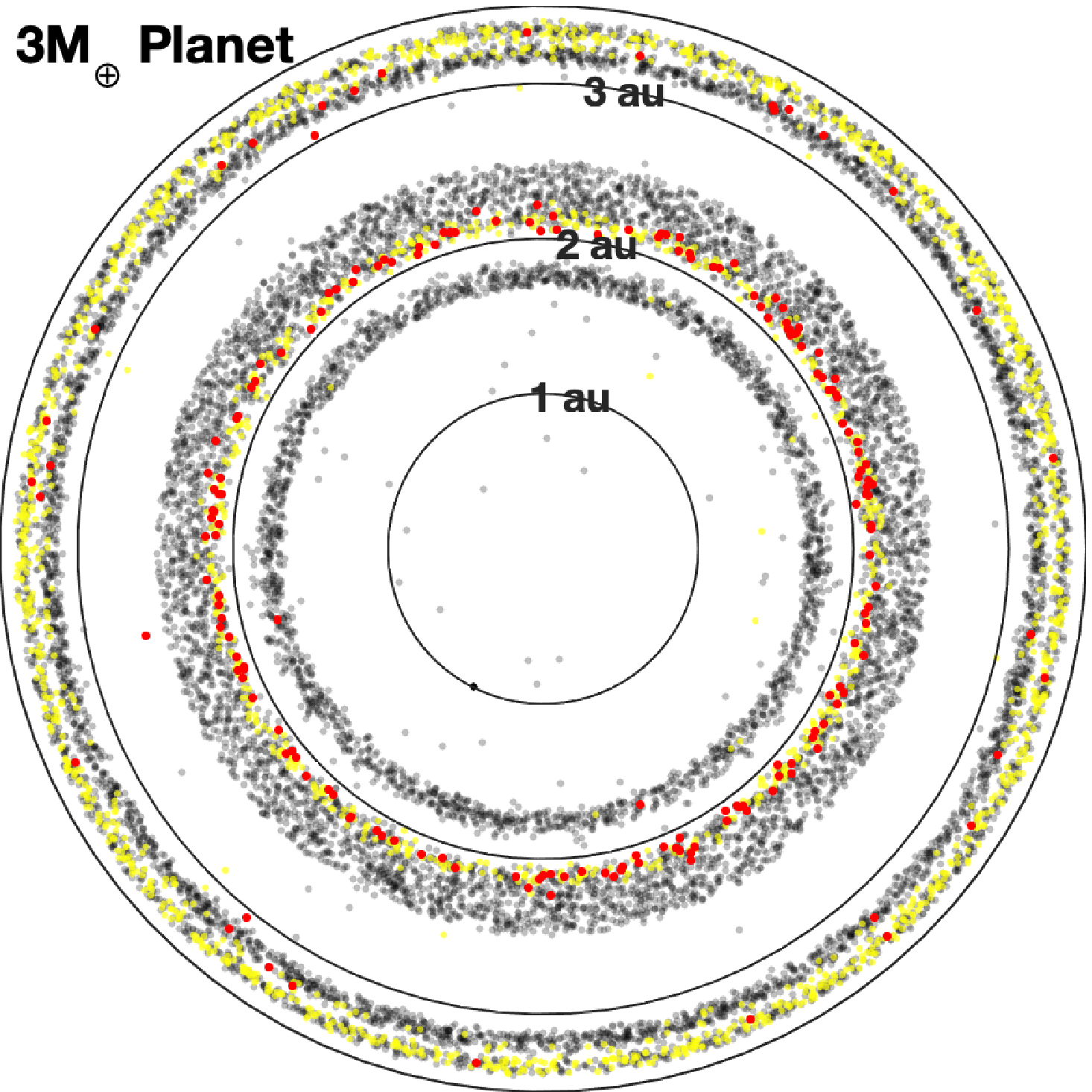}
\caption{Two dimensional polar plots of the particle distributions for asteroids around the $\nu_6$ secular resonance and the 2:1 mean-motion resonance. The orbital phases of the particles are at the beginning of the WD simulations. The upper left plot shows the initial distribution.  The upper right panels denotes the distribution with initially no inner planet. The bottom left and right plots show the distributions with a $1\, \rm M_{\oplus}$ inner planet and a $3\, \rm M_{\oplus}$ inner planet, respectively, after a time of $100\, \rm Myr$. The dot colors denote the outcomes of the particles during the white dwarf stage with yellow denoting ejection, red depicting tidal disruptions, and black indicating stable particles.}
\label{fig::polar}
\end{figure*}

\subsection{Main-sequence stage}
As noted above, to model the shift in the $\nu_6$ secular resonance, we first simulate a belt of test particles during the main-sequence stage for $100\, \rm Myr$. This stage includes the Sun, an inner planet, Jupiter, and Saturn, with two different values of the inner planet mass ($1\, \rm M_{\oplus}$ and $3\, \rm M_{\oplus}$) both at a semi--major axis of $1\,\rm au$. We also explore the relative efficiency of driving tidal disruption events between the $\nu_6$ resonance and the 2:1 mean-motion resonance for each.

 Figure~\ref{fig::ms_v6} shows the number distribution of surviving test particles as a function of semi-major axis at time $t = 0, 5, 10, 50, 100\, \rm Myr$ around the $\nu_6$ secular resonance. Test particles are removed from the simulations through ejections, collisions with the central star or collisions with the planets. The  blue bars correspond  to model  v6\_1ME\_ms  which has a $1\, \rm M_{\oplus}$ inner planet, while the  green bars correspond to model  v6\_3ME\_ms  which has a  $3\, \rm M_{\oplus}$ inner planet.  The gray bars represent the overlap of the two models. The structure of the $\nu_6$ resonance gap remains roughly constant in time after about $50\, \rm Myr$. The timescale of secular perturbations for the $\nu_6$ resonance begins at a time of order $0.1 \, \rm Myr$ \citep{Malhotra1989,Malhotra1999,malhotra2012}.  There is a larger population of removed particles in the $3\, \rm M_{\oplus}$ inner planet simulation compared to the $1\, \rm M_{\oplus}$ inner planet simulation. This was predicted by the secular theory (Fig.~\ref{width}), where the excitation region for the $\nu_6$ resonance is wider for when a $3\, \rm M_{\oplus}$ inner planet is present compared to a $1\, \rm M_{\oplus}$ inner planet. The wider excitation region causes more particles to be cleared from the gap. Both models evolve to close to a steady-state of asteroid clearing within $100\, \rm Myr$,  meaning that the loss rate at $100\, \rm Myr$ has become constant. The final conditions of the $\nu_6$ main-sequence models are taken as the initial conditions for the $\nu_6$ white dwarf models.

Next, we examine the particle distribution around the 2:1 mean-motion resonance in Fig.~\ref{fig::ms_21}. The blue bars correspond to model  21\_1ME\_ms , while the  green bars correspond to model  21\_3ME\_ms.  The gray bars represent the overlap of the two models. The structure of the 2:1 resonance gap remains roughly the same  after about $50\, \rm Myr$. Increasing the mass of the inner planet by a factor of three does not have a significant effect on the dynamics of the 2:1 mean-motion resonance. 

Figure~\ref{fig::e_dist} shows the initial and final distributions of asteroids in semi--major axis and orbital eccentricity of the belt for simulation models  v6\_1ME\_ms, v6\_3ME\_ms, 21\_1ME\_ms, and 21\_3ME\_ms in table~\ref{table::setup}. The top row shows the initial conditions for the belts around the $\nu_6$ secular resonance (left) and the 2:1 mean-motion resonance (right). The middle row shows the distribution of the belts after $100\, \rm Myr$ with a $1\, \rm M_{\oplus}$ inner planet, while the bottom row shows the  distribution of the belts with a $3\, \rm M_{\oplus}$ inner planet.  As expected, the 2:1 resonance gap is qualitatively unaffected by the mass of the inner planet. However, for the $\nu_6$ resonance gap, the simulation with a $3\, \rm M_{\oplus}$ inner planet has a wider gap, and the gap is shifted inward in agreement with the results in Section~\ref{sec:sec_theory} (see Fig.~\ref{width}).  There is also some eccentricity growth at certain mean-motion resonances locations. The 5:1 MMR is located at $a/a_{\rm J} \sim 0.34$ and the 7:2 MMR at $a/a_{\rm J} \sim 0.434$, where there are few tidal disruptions and ejections. Moreover, the 4:1 MMR sits at $a/a_{\rm J} \sim 0.40$ and may affect the nonlinear behaviour of the secular resonance \cite[e.g.,][]{malhotra2012}.

  We set up two additional simulations for the secular resonance and 2:1 mean-motion resonance cases, where the main-sequence stage and the white dwarf stage are simulated with no inner planet in either case. The forced secular eccentricity in the lowest-order theory is independent of the stellar mass, so this is a test of how important higher order terms are in destabilising bodies. Each stellar evolutionary stage is simulated for $100\, \rm Myr$. The bottom left panel of Fig.~\ref{fig::e_dist} shows the eccentricity distribution as a function of semi-major axis of the particles at the end of the main-sequence stage of the $\nu_6$ resonance. By comparing the $\nu_6$ resonance gap structure after $100\, \rm Myr$ to the gap structure produced with the inner planet included, we see that the inner edge of the gap in the "no planet" simulation is shifted outward. There are multiple mean-motion resonances within the gap that have stabilised particles during the main-sequence stage. The bottom right panel of Fig.~\ref{fig::e_dist} shows the eccentricity distribution as a function of semi-major axis of the particles at the end of the main-sequence stage of the 2:1 resonance. The gap size of the 2:1 mean-motion resonance with no inner planet is qualitatively similar to when an inner planet is engulfed.  

To better visualize the belt structure, we show 2-dimensional polar plots in Fig~\ref{fig::polar}.  The points initially interior to $3\, \rm au$ represent particles in the secular resonance simulation and the points initially beyond $3\, \rm au$ represent particles in the 2:1 mean-motion resonance simulation.  The upper left plot represents the initial distribution, while the upper right, bottom left, and bottom right panels show the results with no inner planet, a $1\, \rm M_{\oplus}$ inner planet  and a $3\, \rm M_{\oplus}$ inner planet, respectively. The $\nu_6$ resonance gap is clearly wider and shifted inward when the belt is under the influence of a more massive inner planet.


\subsection{White dwarf stage}
This section describes the results for models  v6\_no\_planet\_wd, v6\_1ME\_wd,  v6\_3ME\_wd,  21\_1ME\_wd, and  21\_3ME\_wd from Table~\ref{table::setup}. As noted, the initial conditions of these models were taken from the final conditions of the main-sequence simulations, models   v6\_1ME\_ms,  v6\_3ME\_ms,  21\_1ME\_ms, and  21\_3ME\_ms, except that the inner planet was removed due to engulfment, the mass of the star was halved, and the orbital radii of all surviving asteroidal bodies  and planets were expanded adiabatically. For the model v6\_no\_planet\_wd, there was no inner planet initially during the main-sequence stage (only Jupiter and Saturn). As discussed earlier, when the inner planet is engulfed, the free precession frequencies of all surviving bodies are altered which causes the $\nu_6$ secular resonance to shift outwards.  We instantaneously remove the inner planet and reduce the stellar mass. In practise, the inner planet's orbit may decay through tidal decay before the plant is engulfed, This will cause the secular resonances to move smoothly through the system \cite[see fig. 14 of][which only dealt with exterior secular resonances]{Mustill2012}.  The effects of this are slightly different from the impulsive change in location in this study, particularly if the resonance moves a substantial distance. Nevertheless, the dynamics of the secular resonance present in this work will have profound effects on the study of white dwarf pollution.

The left panels of Fig.~\ref{fig::e_dist} shows the initial particle eccentricities as a function of initial semi-major axes for the $1\, \rm M_{\oplus}$ engulfed planet (middle panel) and the $3\, \rm M_{\oplus}$ engulfed planet (bottom panel) simulations, models  v6\_1ME\_wd and  v6\_1ME\_wd, respectively. The yellow colour denotes particles that have been ejected  after the star became a white dwarf, and red denotes particles that have been tidally disrupted by the white dwarf. The black colour represents particles that have remained stable throughout the simulation time domain. As shown in the figure, the majority of ejections/tidal disruptions occur as a result of the outward shift of the $\nu_6$ secular resonance.

 With no inner planet engulfment, i.e. the "no planet" simulation, we still see the secular resonance shifts outward in Fig~\ref{fig::e_dist}. However, by comparing the $\nu_6$  resonance gap structure after $100\, \rm Myr$ of the "no planet" simulation to the gap structure with a $1\, \rm M_{\oplus}$ planet, we see that the inner edge of the gap in the "no planet" simulation is shifted outward. This means that the gap is more centered on the neighboring mean-motion resonances. When a secular resonance is overlapping with  mean-motion renounces, the destabilising perturbations are exacerbated \cite[e.g.,][]{moons1995}.  Therefore, with no inner planet engulfment the secular resonance still shifts outward. The forced secular resonance eccentricity in the higher-order theory does have a significant effect on the secular perturbations once the system evolves to a white dwarf. The significance of this result, is that  white dwarf pollution by secular resonances is much more robust since inner planetary engulfment is not necessarily required. Figure~\ref{fig::e_dist} also shows a similar gap structure of the 2:1 resonance without having an inner planet as it does with the inclusion of the inner planet (right panels of Fig.~\ref{fig::e_dist}).

 The right panels of Fig.~\ref{fig::e_dist} shows
 similar plots for the particles around the 2:1 mean-motion resonance. Due to the mass loss from the central star, the mean-motion resonance width increases and causes destabilization of particles that were once stable during the main-sequence stage \citep{Debes2012}. The yellow colour denotes particles that have been ejected, and red denotes particles that have been tidally disrupted by the white dwarf. The majority of the outcomes are ejections rather than tidal disruptions. The locations of particles that have been ejected or tidally disrupted around both the 2:1 mean-motion resonance and the secular resonance can be seen in the bottom and right plots in Fig.~\ref{fig::polar}.

  \begin{figure}
\includegraphics[width=\columnwidth]{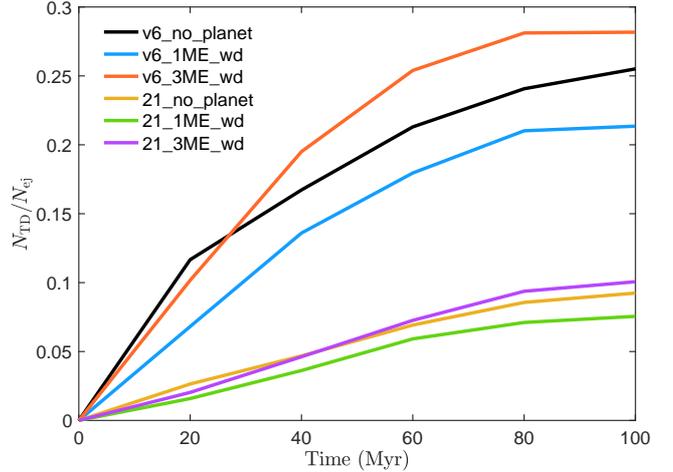}
\caption{  The ratio of the number of tidally disrupted particles ($N_{\rm TD}$) to the number of ejected particles ($N_{\rm ej}$) as a function of time for models v6\_no\_planet (black), v6\_1ME\_wd (blue), v6\_3ME\_wd (red), 21\_no\_planet (yellow), 21\_1ME\_wd (green), and 21\_3ME\_wd (purple). See Table~\ref{table::setup} for a description of the different models.
}
\label{fig::num_events_POST}
\end{figure}

\begin{figure}
\includegraphics[width=\columnwidth]{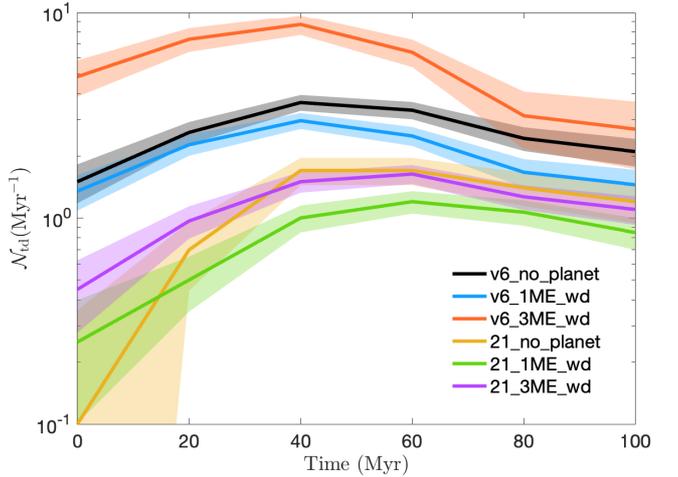}
\caption{ The number of tidally disrupted particles per Myr, $\mathcal{N}_{\rm td }$, as a function of time. The black, blue, and red lines denote the secular resonance models (v6\_no\_planet, v6\_1ME\_wd, and v6\_3ME\_wd; see Table~\ref{table::setup}), with no inner planet, a $1\, \rm M_{\oplus}$, and a $3\, \rm M_{\oplus}$ engulfed inner planet, respectively. The yellow, green, and purple lines denote the 2:1 mean-motion resonance models (21\_no\_planet, 21\_1ME\_wd, and 21\_3ME\_wd). The shaded regions identify the standard error in the tidal disruption rate for each model.
}
\label{fig::td_rate}
\end{figure}

 Figure~\ref{fig::num_events_POST} shows the ratio of the number of tidally disrupted particles ($N_{\rm TD}$) to the number of ejected particles ($N_{\rm ej}$) for both the secular resonance models and the 2:1 mean-motion resonance models.  For each model, the $N_{\rm TD}/N_{\rm ej}$ ratio gradually increases with time. The ratio increases more substantially for the secular resonance models compared to the mean motion resonance. In each case the ratio is larger when a $3\, \rm M_{\oplus}$ mass planet is engulfed compared to when a $1\, \rm M_{\oplus}$ mass planet is engulfed. However, the difference in the ratio between the two engulfed planet masses is more significant for the secular resonance models than the mean-motion models because increasing the mass of the engulfment planet does not have much impact on the dynamics of mean-motion resonances. For the secular resonance models, this difference arises because the resonance orbital location for the higher mass planet is initially closer-in.  The "no planet" simulation mapping the $\nu_6$ resonance has a higher  $N_{\rm TD}/N_{\rm ej}$ ratio than the $1\, \rm M_{\oplus}$ mass planet engulfment scenario. The $N_{\rm TD}/N_{\rm ej}$ ratio for the "no planet" simulation mapping the 2:1 resonance is roughly the same as when the inner planet is included. 
 
 These ratios for the 2:1 mean-motion resonance models are relatively small due to that fact that there are significantly more ejections than tidal disruptions compared to the secular resonance case. The 2:1 resonance has a higher probability of producing ejections rather than tidal disruptions because it has a larger orbital radius than the $\nu_6$ secular resonance. Note that even for the $\nu_6$ resonance case, there are more ejections than tidal disruptions. The fact that the ratios for the 2:1 mean-motion resonance models are lower than the ratios from the secular resonance models means that the secular resonance is more efficient in driving tidal disruption events.  On a more fundamental level, the $\nu_6$ secular resonance is more efficient in driving white dwarf pollution than {\it all} mean-motion resonances, since the 2:1 mean-motion resonance is a first-order resonance, and hence the strongest resonance.

 Next we compare the absolute rates of tidally disrupted particles between the secular and mean-motion resonance models.  Note that when comparing the tidal disruption rates between the different models, we use the same number of particles per semi-major axis when populating the two resonances. To calculate the tidal disruption rate for a real system, one would have to apply a weighting given the belt surface density profile. Figure~\ref{fig::td_rate} shows the number of tidally disrupted particles per Myr, $\mathcal{N}_{\rm td }$, as a function of time for the white dwarf models  v6\_no\_planet\_wd, v6\_1ME\_wd, v6\_3ME\_wd, 21\_no\_planet\_wd, 21\_1ME\_wd, and 21\_3ME\_wd. The first three describe the secular resonance models  (including the control simulation), and the last three represent the 2:1 mean-motion resonance models.  The shaded regions identify the standard error in the tidal disruption rate for each model. The secular resonance simulation that assumed a $3\,\rm M_{\oplus}$ engulfed planet ( model v6\_3ME\_wd) has a higher rate of tidally disputed particles  than the simulation that assumed a $1\,\rm M_{\oplus}$ engulfed planet ( model v6\_1ME\_wd).  When no engulfed innner planet is simulated, the rate of disruptions is higher than the $1\,\rm M_{\oplus}$ engulfed planet scenerio, which means that planetary engulfment is not necessarily needed. These two models have overlapping standard error near the beginning of the simulations but then deviate beyond $40\, \rm Myr$. \cite{smallwood2018b} showed that a $1M_\oplus$ mass engulfed inner planet at $1\, \rm au$ can produce a high enough tidal disruption rate to have a mass accretion rate that is within the range of accretion rates deduced from observations, from $\sim 10^5\, \rm g/s$ to  $\sim 10^{11}\, \rm g/s$ \citep{Koester2014,Farihi2016}.  The tidal disruption rate is estimated to be $0.0002\, \rm Myr^{-1}$ per particle for a fiducial belt mass. Thus, the tidal disruption rates for each secular resonance model are able to provide a mass accretion rate that is within the observed limits. The rates of tidally disrupted particles for the 2:1 mean-motion simulations,  21\_no\_planet\_wd, 21\_1ME\_wd and 21\_3ME\_wd, are lower than the tidal disruption rates for the secular resonance models, which again suggests that the secular resonance is more efficient in producing the required accretion rates than the 2:1 mean-motion resonance. Still, both mean-motion resonances and secular resonances are expected to contribute to the accretion process.

\section{Known Exoplanetary Systems}
\label{sec:exoplanet}

White dwarf pollution may occur in planetary systems that are very different from our solar system. Secular and mean-motion resonances are expected to sculpt the architecture of an asteroid belt (if one exists) in exoplanetary systems. Consequently, the mechanisms presented in this work should occur in numerous planetary configurations.  In this Section we consider the potential for white dwarf pollution in observed exoplanetary systems due to planetary engulfment.

From the {\it Kepler} data,  planetary systems are common with  an  occurrence rate of at minimum one planet per star \citep{Petigura2013,ForemanMackey2014,Burke2015,Hsu2018}.  This occurrence rate is also seen from microlensing observations \citep{Cassan2012}.  There are, however, strong observational biases that are inherent in transit surveys \cite[e.g.,][]{Kipping2016}. Small planets with long periods are much more difficult to detect than giant planets orbiting nearby to their star.  This is because of the transit signal-to-noise which prevents the detection of small planets with periods $\sim 1\, \rm yr$. Due to the operational time of {\it Kepler} and a planet confirmation criterion of three transits, only giant planets were detectable with periods up to $418$ days \citep{Fressin2013}. Large self-luminous planets are more sensitive to direct imaging at large separation, $\gtrsim 10\, \rm au$ \citep{Kalas2008,Marois2008,Marois2010,Lagrange2010}. The planet detections to date suggest that giant planets are more common around A stars and that wide-separation planets are more prevalent around high-mass stars \citep{Johnson2010,Reffert2014}.  Furthermore, the occurrence rate of giant planets around stars from  direct imaging statistics is on the order of $10\%$ \citep{Galicher2016,Meshkat2017,Nielsen2019,Baron2019}. Moreover, \cite{Wahhaj2013} found, from the Gemini NICI Planet-Finding Campaign, that $<20\%$ of debris disk stars have a $\leq 3\, \rm M_{Jup}$ planet beyond $10\, \rm au$.

\begin{figure}
\includegraphics[width=\columnwidth]{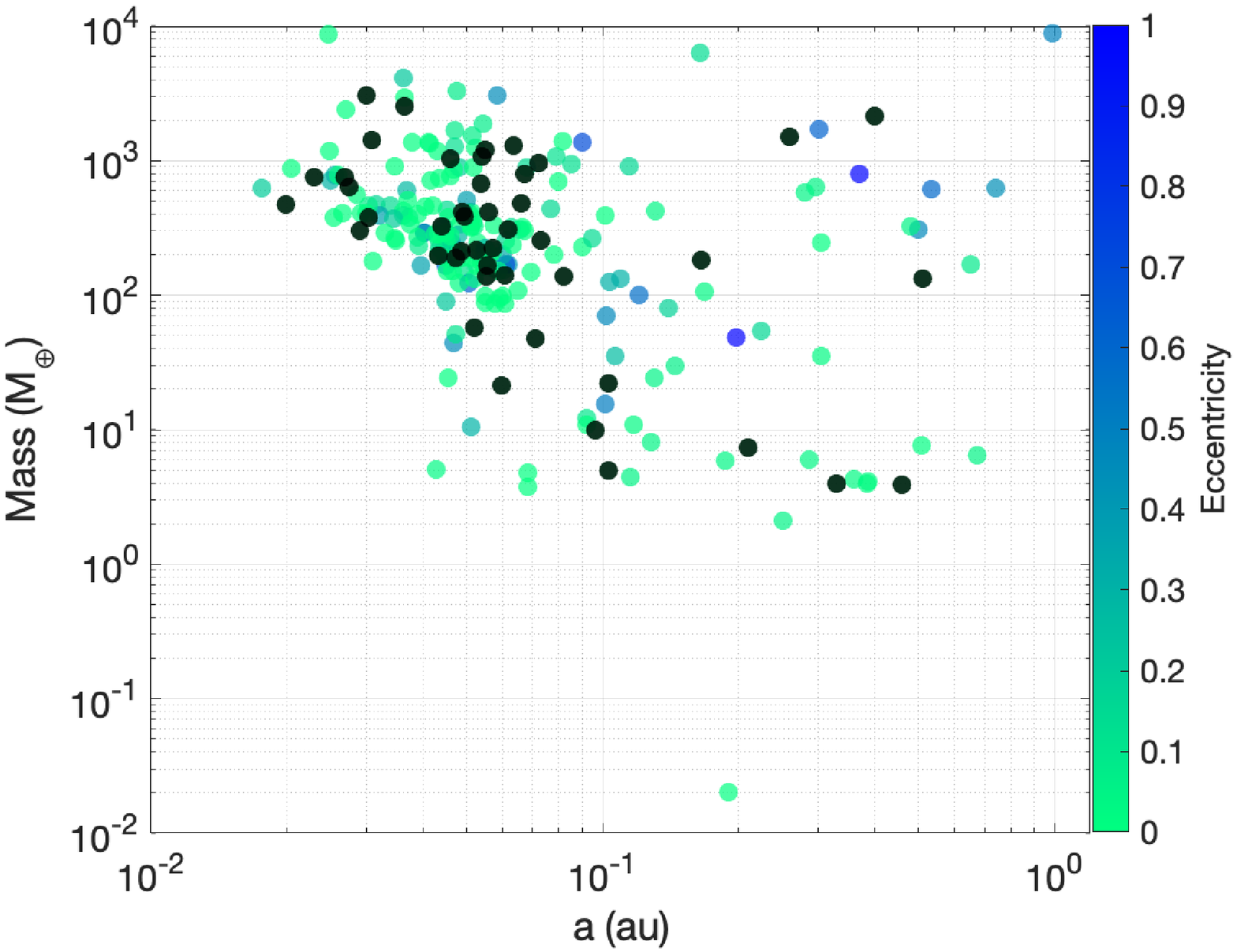}
\includegraphics[width=\columnwidth]{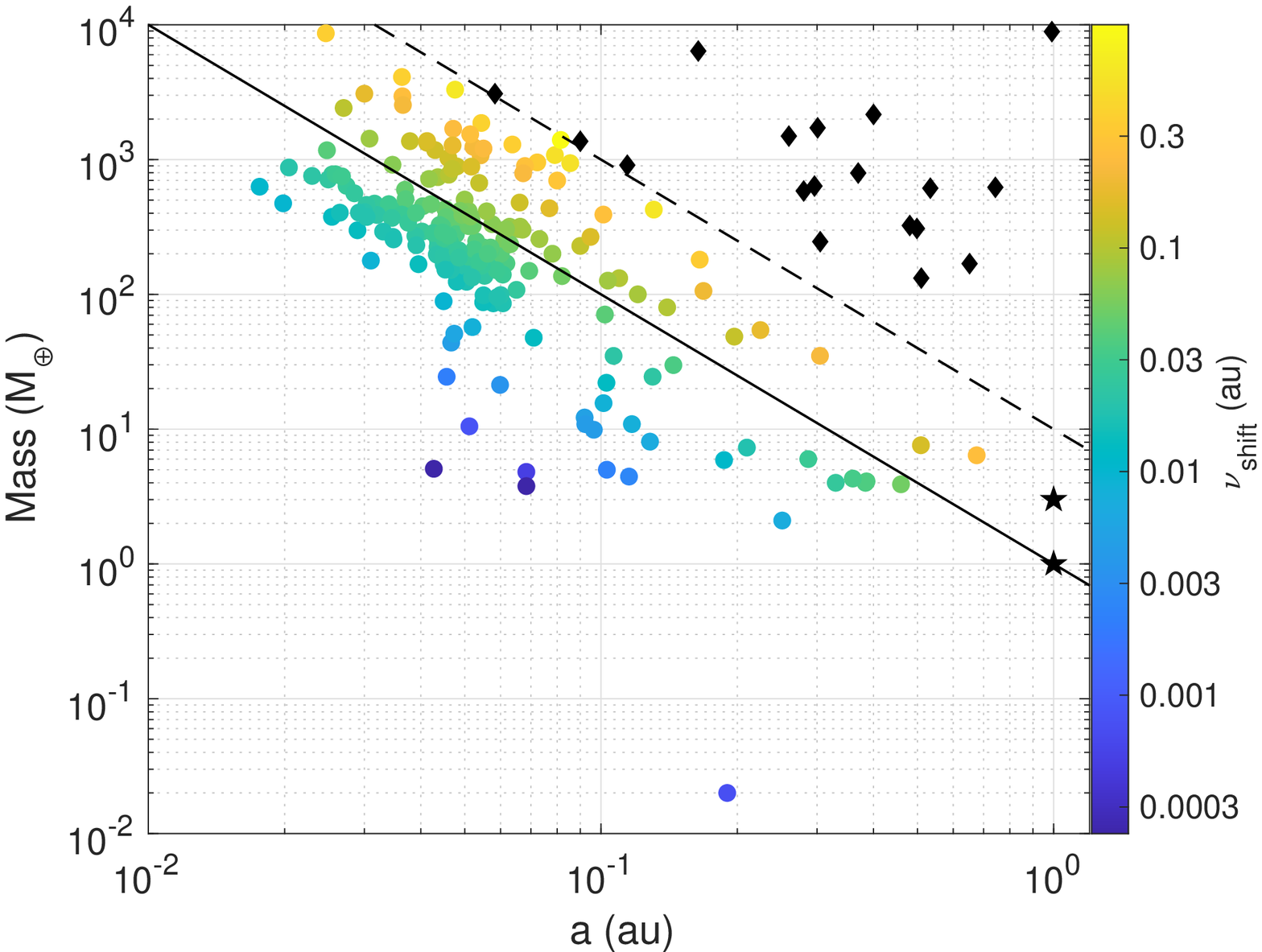}
\caption{Upper panel: The mass versus semi-major axis for  exoplanets that have a host star with mass in a range that will become a white dwarf. The eccentricity of each exoplanet is denoted by the color bar. The black dots are exoplanets that have unknown eccentricity. Bottom panel: Same as the upper panel but the color bar now denotes the secular resonance shift distance ($\nu_{\rm shift}$) once the planet  has been engulfed. The diamond markers indicate the engulfed planets that cause a newly formed secular resonance. The two star markers represent a $1M_{\oplus}$ inner planet and a $3M_{\oplus}$ inner planet located at $1\, \rm au$, as used in the numerical simulations in Section~\ref{sec:Nbody}. Systems above the solid line have a resonance shift $\gtrsim 0.06\, \rm au$ while those above the dashed line do not have a resonance before the white dwarf stage.
}
\label{exo_data_new}
\end{figure}

We apply the secular theory model to exoplanetary systems to study the dynamics of secular resonance shifts due to planetary engulfment. 
For the inner system architecture, we extract the data of all available exoplanetary systems from the NASA exoplanet archive\footnote{https://exoplanetarchive.ipac.caltech.edu}. We filter the exoplanets to those that (1) have the host star within a mass range that would produce a white dwarf, $1.0 M_{\odot} \lesssim M \lesssim 10 M_{\odot}$,  lower-mass stars have not had time to become white dwarfs, (2) have a planet with a known mass, and (3) a semi-major axis $\leq 1\, \rm au$ so that they will be engulfed.  The two variables that dominate the secular theory calculations are the planet mass and semi-major axis, thus if the eccentricity or argument of the pericenter is unknown, we assume values of $0.0$ and $90\degree$, respectively. There are $230$ exoplanets that fit these criteria. We show the mass as a function of the semi-major axis along with the eccentricity for each exoplanet in the upper panel of Fig.~\ref{exo_data_new}. The black dots are the exoplanets that have an unknown eccentricity.

 For each exoplanet in our sample, we assume that its outer system architecture has giants like Jupiter and Saturn  (which are difficult to detect but likely to exist, even though not with those exact parameters) and we then measure the location of the secular resonance before planetary engulfment and then compare to the location after inner planet engulfment.

 The bottom panel of Fig.~\ref{exo_data_new} shows the mass semi-major axis distribution of the exoplanets along with the secular resonance shift distance ($\nu_{\rm shift}$) denoted by the color. The $1\, \rm M_{\oplus}$ and $3\, \rm M_{\oplus}$ inner planets used in the $N$-body simulations are shown by the star markers. There are systems that have sufficiently massive inner planets that there would currently be no resonance  because the asteroid free precession rate is larger than the proper mode of Saturn within the asteroid belt region. The engulfment of these planets leads to the formation of a secular resonance (shown by the diamond markers) and the maximum number of tidal disruption events possible occurs for these planets.  The dashed black line shows the approximate boundary between forming a new resonance and having a resonance shift after engulfment. The line is given by $ M=10\left(a/{\rm au}\right)^{-2}\,\rm M_\oplus$.

An Earth mass planet at $1\,\rm au$ is close to the critical mass and separation for a system in which the resonance moves more than its width and thus a significant number of tidal disruption events occurs.  We approximate the mass required for a significant shift in the resonance by 
$ M>\left(a/{\rm au}\right)^{-2}\,\rm M_\oplus$.
We plot this solid black line in the lower panel of Fig.~\ref{exo_data_new}.  An Earth mass planet at a smaller orbital radius or a smaller mass planet at the same location causes a smaller resonance shift.

In the solar neighborhood it has been observed that over half of the Sun-like stars have at least one super-Earth planet orbiting on a low eccentricity orbit with a period of days to months \citep{Mayor2011,Batalha2013,Fressin2013,Burke2015,MartinLivio2016}. 
A planetary system with an engulfed super-Earth at $1\, \rm au$ will cause a resonance shift of $> 0.06 \,\rm au$ if the mass is $1 < M/M_{\oplus} < 10$.  The value of $0.06\, \rm au$ is the resonance shift distance from an engulfed $ 1\, \rm  M_{\oplus}$ planet. However, if the super--Earth mass is $\geq 10\,\rm M_\oplus$, then there will not be a secular resonance present until the super-Earth is engulfed \citep{smallwood2018a},  which leads to the formation of the resonance after the planet is engulfed and thus the most substantial rate of tidal disruption events. A super--Earth with a mass of $10\,\rm M_{\oplus}$ could have a semi--major axis as low as $0.31\,\rm au$ and cause a shift in the resonance of $0.06\,\rm au$.


 Hot Jupiters orbit their central star at distances of $\lesssim 0.05\, \rm au$ \citep{Hartman2012,Hellier2012,Weiss2013}. If  a hot Jupiter at a semi-major axis of $0.05\, \rm au$ is accompanied by two outer giant planets, we can constrain the mass required for the engulfed hot Jupiter to produce a significant resonance shift to be
about $>1.3\, \rm M_J$.  However, hot Jupiters rarely dwell in multi-planet systems \citep{Wright2009,Gibson2009,Latham2011,Steffen2012}.  There are, though, several exceptions, with one being the WASP-47 system which hosts a hot Jupiter along with a Neptune-sized outer companion, a super-Earth inner planet, and a Jupiter-sized planet with a separation of $1\, \rm au$ \citep{Becker2015,NeveuVanMalle2016,Weiss2017}. Other systems that host a hot Jupiter along with outer giant companions include: HIP 14810 \citep{Ment2018}, HD 217107 \citep{Wright2009,Stassum2017}, Pr0211 \citep{Malavolta2016}, and HD 187123 \citep{Wright2009}. The lack of observed companions in hot Jupiter systems can be caused by low sensitively in transit timing variations \citep{Steffen2012} or large mutual inclinations of the companions \citep{Triaud2010,Winn2010,Morton2011,Zhou2015}. In conclusion, engulfing a hot Jupiter can cause a white dwarf to become polluted if there are two outer giant companions present.  It should be noted that hot Jupiters would be engulfed very soon after the star leaves the main sequence, after which it takes $\sim10^8-10^9$ years (depending on mass) to become a white dwarf. There is therefore the risk that the new location of the secular resonance gets depleted while the star is a giant, leaving little material to survive to pollute the white dwarf.

The results in this section depend on there being two outer massive planets (i.e., Jupiter and Saturn). By varying the properties of these two planets, the location and resulting shift of the secular resonance will be different. For a $1\, \rm M_{\oplus}$ inner engulfed planet at $1\, \rm au$, changing the mass of Jupiter by a factor of two will cause the resonance shift distance to increase by a factor of about $3$, which would cause the area between the two lines in the bottom panel of Fig.~\ref{exo_data_new} to expand. Likewise, by changing the mass of Saturn by a factor of two, the resonance would shift inward rather than outward \citep{smallwood2018a}. Thus, for this mechanism to produce tidal disruption events of asteroidal bodies, the combination of the parameters of the engulfed planet and the two outer giant planets must work synergistically.


\section{Conclusions}
\label{sec:conc}

In \cite{smallwood2018b} we found that the $\nu_6$ secular resonance in the solar system shifts outward when the Earth is engulfed during stellar evolution. This resonance shift causes previously stable asteroids to undergo secular resonant perturbations, to move to star-grazing orbits and become tidally disrupted by the white dwarf. In this work, we investigated how the resonance shift is related to the mass and semi-major axis of the inner engulfed planet. From secular analytical theory, we found that the more massive the engulfed planet and the larger its orbital separation, the more significant is the resonance shift and the higher the rate of tidal disruption events of asteroids.  Hence, if secular resonance sweeping is the dominant mechanism to pollute white dwarfs, then the currently-observed rate of pollution may represent a proxy for the engulfment history of the star.

We ran higher order numerical simulations with two different inner planet masses, $M = 1$ and $3\, \rm M_{\oplus}$ at a fixed semi-major axis of $1\,\rm au$, to validate our second-order secular theory results. We found that the secular resonance does shift outward as found by \cite{smallwood2018b}. Furthermore, when a $3\, \rm M_{\oplus}$ planet is engulfed, the number of tidally disrupted particles increases (when compared to a lower mass engulfed planet). The tidal disruption rate for both secular resonance models can give an accretion rate in agreement with those deduced from pollution data.

 Moreover, we now find that a planet does not necessarily have to be engulfed in order to trigger secular resonant perturbations during the white dwarf stage. Our "no planet" simulation, which includes no inner planetary engulfment, shows that higher-order terms in the secular theory (not captured by our analytical calculations) are important in destablising particles near secular resonances. However, a more massive engulfed planet will still have a higher tidal disruption rate.

We also tested the efficiency of driving tidal disruption events during the white dwarf stage for particles around the 2:1 mean-motion resonance, which is one of, if not the strongest mean-motion resonances. The tidal disruption rate from the 2:1 mean-motion resonance is lower than that from the secular resonance, which demonstrates that the  $\nu_6$ resonance is more efficient in driving white dwarf pollution than mean-motion resonances.

Finally, we explored the feasibility of a secular resonance shift within observed exoplanetary systems. 
We assumed an architecture similar to that of  the outer Solar System (i.e., Jupiter and Saturn) and estimated the secular resonance shift by engulfing known exoplanets with a semi-major axis $\leq 1\, \rm au$. We found that in addition to an Earth mass planet at $1\,\rm au$,  hot Jupiters very close to their star and super-Earths farther out are able to produce similar tidal disruption rates.  Thus, the mechanism of white dwarf pollution through secular resonances appears to be robust since planetary engulfment is not necessarily required and should operate for a significant fraction of  the observed exoplanetary inner system architectures.

\section*{Acknowledgements}
 We much appreciate Alexander Mustill  for carefully reviewing the paper. We thank Steve Lubow for insightful discussions. Computer support was provided by UNLV's National Supercomputing Center. This research has made use of data and/or services provided by the NASA exoplanet archive.  RGM acknowledges support from NASA through grant NNX17AB96G. DV gratefully acknowledges the support of the STFC via an Ernest Rutherford Fellowship (grant ST/P003850/1).

\section*{Data Availability}

The data supporting the plots within this article are available on reasonable request to the corresponding author. A public version of the {\sc mercury} code is available at \url{https://github.com/4xxi/mercury}.




\bibliographystyle{mnras}
\bibliography{smallwood} 

\begin{thebibliography}{}
\makeatletter
\relax
\def\mn@urlcharsother{\let\do\@makeother \do\$\do\&\do\#\do\^\do\_\do\%\do\~}
\def\mn@doi{\begingroup\mn@urlcharsother \@ifnextchar [ {\mn@doi@}
  {\mn@doi@[]}}
\def\mn@doi@[#1]#2{\def\@tempa{#1}\ifx\@tempa\@empty \href
  {http://dx.doi.org/#2} {doi:#2}\else \href {http://dx.doi.org/#2} {#1}\fi
  \endgroup}
\def\mn@eprint#1#2{\mn@eprint@#1:#2::\@nil}
\def\mn@eprint@arXiv#1{\href {http://arxiv.org/abs/#1} {{\tt arXiv:#1}}}
\def\mn@eprint@dblp#1{\href {http://dblp.uni-trier.de/rec/bibtex/#1.xml}
  {dblp:#1}}
\def\mn@eprint@#1:#2:#3:#4\@nil{\def\@tempa {#1}\def\@tempb {#2}\def\@tempc
  {#3}\ifx \@tempc \@empty \let \@tempc \@tempb \let \@tempb \@tempa \fi \ifx
  \@tempb \@empty \def\@tempb {arXiv}\fi \@ifundefined
  {mn@eprint@\@tempb}{\@tempb:\@tempc}{\expandafter \expandafter \csname
  mn@eprint@\@tempb\endcsname \expandafter{\@tempc}}}

\bibitem[\protect\citeauthoryear{{Aannestad}, {Kenyon}, {Hammond}  \&
  {Sion}}{{Aannestad} et~al.}{1993}]{Aannestad1993}
{Aannestad} P.~A.,  {Kenyon} S.~J.,  {Hammond} G.~L.,   {Sion} E.~M.,  1993,
  \mn@doi [\aj] {10.1086/116491}, \href
  {http://adsabs.harvard.edu/abs/1993AJ....105.1033A} {105, 1033}

\bibitem[\protect\citeauthoryear{{Adams} \& {Bloch}}{{Adams} \&
  {Bloch}}{2013}]{Adams2013}
{Adams} F.~C.,  {Bloch} A.~M.,  2013, \mn@doi [\apjl]
  {10.1088/2041-8205/777/2/L30}, \href
  {http://adsabs.harvard.edu/abs/2013ApJ...777L..30A} {777, L30}

\bibitem[\protect\citeauthoryear{{Althaus}, {C{\'o}rsico}, {Isern}  \&
  {Garc{\'{\i}}a-Berro}}{{Althaus} et~al.}{2010}]{Althaus2010}
{Althaus} L.~G.,  {C{\'o}rsico} A.~H.,  {Isern} J.,   {Garc{\'{\i}}a-Berro} E.,
   2010, \mn@doi [\aapr] {10.1007/s00159-010-0033-1}, \href
  {http://adsabs.harvard.edu/abs/2010A%26ARv..18..471A} {18, 471}

\bibitem[\protect\citeauthoryear{{Baron}, {Lafreni{\`e}re}, {Artigau},
  {Gagn{\'e}}, {Rameau}, {Delorme}  \& {Naud}}{{Baron}
  et~al.}{2019}]{Baron2019}
{Baron} F.,  {Lafreni{\`e}re} D.,  {Artigau} {\'E}.,  {Gagn{\'e}} J.,  {Rameau}
  J.,  {Delorme} P.,   {Naud} M.-E.,  2019, \mn@doi [\aj]
  {10.3847/1538-3881/ab4130}, \href
  {https://ui.adsabs.harvard.edu/abs/2019AJ....158..187B} {158, 187}

\bibitem[\protect\citeauthoryear{{Barstow}, {Barstow}, {Casewell}, {Holberg}
  \& {Hubeny}}{{Barstow} et~al.}{2014}]{Barstow2014}
{Barstow} M.~A.,  {Barstow} J.~K.,  {Casewell} S.~L.,  {Holberg} J.~B.,
  {Hubeny} I.,  2014, \mn@doi [\mnras] {10.1093/mnras/stu216}, \href
  {http://adsabs.harvard.edu/abs/2014MNRAS.440.1607B} {440, 1607}

\bibitem[\protect\citeauthoryear{{Batalha} et~al.,}{{Batalha}
  et~al.}{2013}]{Batalha2013}
{Batalha} N.~M.,  et~al., 2013, \mn@doi [\apjs] {10.1088/0067-0049/204/2/24},
  \href {http://adsabs.harvard.edu/abs/2013ApJS..204...24B} {204, 24}

\bibitem[\protect\citeauthoryear{{Bauer} \& {Bildsten}}{{Bauer} \&
  {Bildsten}}{2019}]{Bauer2019}
{Bauer} E.~B.,  {Bildsten} L.,  2019, \mn@doi [\apj]
  {10.3847/1538-4357/ab0028}, \href
  {https://ui.adsabs.harvard.edu/abs/2019ApJ...872...96B} {872, 96}

\bibitem[\protect\citeauthoryear{{Becker}, {Vanderburg}, {Adams}, {Rappaport}
  \& {Schwengeler}}{{Becker} et~al.}{2015}]{Becker2015}
{Becker} J.~C.,  {Vanderburg} A.,  {Adams} F.~C.,  {Rappaport} S.~A.,
  {Schwengeler} H.~M.,  2015, \mn@doi [\apjl] {10.1088/2041-8205/812/2/L18},
  \href {https://ui.adsabs.harvard.edu/abs/2015ApJ...812L..18B} {812, L18}

\bibitem[\protect\citeauthoryear{{Becklin}, {Farihi}, {Jura}, {Song},
  {Weinberger}  \& {Zuckerman}}{{Becklin} et~al.}{2005}]{Becklin2005}
{Becklin} E.~E.,  {Farihi} J.,  {Jura} M.,  {Song} I.,  {Weinberger} A.~J.,
  {Zuckerman} B.,  2005, \mn@doi [\apjl] {10.1086/497826}, \href
  {https://ui.adsabs.harvard.edu/abs/2005ApJ...632L.119B} {632, L119}

\bibitem[\protect\citeauthoryear{{Blouin}}{{Blouin}}{2020}]{Blouin2020}
{Blouin} S.,  2020, \mn@doi [\mnras] {10.1093/mnras/staa1689}, \href
  {https://ui.adsabs.harvard.edu/abs/2020MNRAS.496.1881B} {496, 1881}

\bibitem[\protect\citeauthoryear{{Bonsor} \& {Veras}}{{Bonsor} \&
  {Veras}}{2015}]{Bonsor2015}
{Bonsor} A.,  {Veras} D.,  2015, \mn@doi [\mnras] {10.1093/mnras/stv1913},
  \href {http://adsabs.harvard.edu/abs/2015MNRAS.454...53B} {454, 53}

\bibitem[\protect\citeauthoryear{{Bonsor}, {Farihi}, {Wyatt}  \& {van
  Lieshout}}{{Bonsor} et~al.}{2017}]{Bonsor2017}
{Bonsor} A.,  {Farihi} J.,  {Wyatt} M.~C.,   {van Lieshout} R.,  2017, \mn@doi
  [\mnras] {10.1093/mnras/stx425}, \href
  {http://adsabs.harvard.edu/abs/2017MNRAS.468..154B} {468, 154}

\bibitem[\protect\citeauthoryear{{Bonsor}, {Carter}, {Hollands},
  {G{\"a}nsicke}, {Leinhardt}  \& {Harrison}}{{Bonsor}
  et~al.}{2020}]{Bonsor2020}
{Bonsor} A.,  {Carter} P.~J.,  {Hollands} M.,  {G{\"a}nsicke} B.~T.,
  {Leinhardt} Z.,   {Harrison} J. H.~D.,  2020, \mn@doi [\mnras]
  {10.1093/mnras/stz3603}, \href
  {https://ui.adsabs.harvard.edu/abs/2020MNRAS.492.2683B} {492, 2683}

\bibitem[\protect\citeauthoryear{{Bromley} \& {Kenyon}}{{Bromley} \&
  {Kenyon}}{2019}]{Bromley2019}
{Bromley} B.~C.,  {Kenyon} S.~J.,  2019, \mn@doi [\apj]
  {10.3847/1538-4357/ab12e9}, \href
  {https://ui.adsabs.harvard.edu/abs/2019ApJ...876...17B} {876, 17}

\bibitem[\protect\citeauthoryear{{Brown}, {Veras}  \& {G{\"a}nsicke}}{{Brown}
  et~al.}{2017}]{Brown2017}
{Brown} J.~C.,  {Veras} D.,   {G{\"a}nsicke} B.~T.,  2017, \mn@doi [\mnras]
  {10.1093/mnras/stx428}, \href
  {http://adsabs.harvard.edu/abs/2017MNRAS.468.1575B} {468, 1575}

\bibitem[\protect\citeauthoryear{{Burke} et~al.,}{{Burke}
  et~al.}{2015}]{Burke2015}
{Burke} C.~J.,  et~al., 2015, \mn@doi [\apj] {10.1088/0004-637X/809/1/8}, \href
  {https://ui.adsabs.harvard.edu/abs/2015ApJ...809....8B} {809, 8}

\bibitem[\protect\citeauthoryear{{Caiazzo} \& {Heyl}}{{Caiazzo} \&
  {Heyl}}{2017}]{Caiazzo2017}
{Caiazzo} I.,  {Heyl} J.~S.,  2017, preprint, \href
  {http://adsabs.harvard.edu/abs/2017arXiv170207682C} {} (\mn@eprint {arXiv}
  {1702.07682})

\bibitem[\protect\citeauthoryear{{Cassan} et~al.,}{{Cassan}
  et~al.}{2012}]{Cassan2012}
{Cassan} A.,  et~al., 2012, \mn@doi [Nature] {10.1038/nature10684}, \href
  {https://ui.adsabs.harvard.edu/abs/2012Natur.481..167C} {481, 167}

\bibitem[\protect\citeauthoryear{{Chambers}}{{Chambers}}{1999}]{Chambers1999}
{Chambers} J.~E.,  1999, \mn@doi [\mnras] {10.1046/j.1365-8711.1999.02379.x},
  \href {http://adsabs.harvard.edu/abs/1999MNRAS.304..793C} {304, 793}

\bibitem[\protect\citeauthoryear{{Cummings}, {Kalirai}, {Tremblay},
  {Ramirez-Ruiz}  \& {Choi}}{{Cummings} et~al.}{2018}]{Cummings2018}
{Cummings} J.~D.,  {Kalirai} J.~S.,  {Tremblay} P.~E.,  {Ramirez-Ruiz} E.,
  {Choi} J.,  2018, \mn@doi [\apj] {10.3847/1538-4357/aadfd6}, \href
  {https://ui.adsabs.harvard.edu/abs/2018ApJ...866...21C} {866, 21}

\bibitem[\protect\citeauthoryear{{Cunningham et al.}}{{Cunningham et
  al.}}{2021}]{Cunningham2021}
{Cunningham et al.} 2021, \mnras, Submitted

\bibitem[\protect\citeauthoryear{{Debes}}{{Debes}}{2006}]{Debes2006}
{Debes} J.~H.,  2006, \mn@doi [The Astrophysical Journal] {10.1086/508132},
  \href {https://ui.adsabs.harvard.edu/abs/2006ApJ...652..636D} {652, 636}

\bibitem[\protect\citeauthoryear{{Debes} \& {Sigurdsson}}{{Debes} \&
  {Sigurdsson}}{2002}]{Debes2002}
{Debes} J.~H.,  {Sigurdsson} S.,  2002, \mn@doi [\apj] {10.1086/340291}, \href
  {http://adsabs.harvard.edu/abs/2002ApJ...572..556D} {572, 556}

\bibitem[\protect\citeauthoryear{{Debes}, {Walsh}  \& {Stark}}{{Debes}
  et~al.}{2012}]{Debes2012}
{Debes} J.~H.,  {Walsh} K.~J.,   {Stark} C.,  2012, \mn@doi [\apj]
  {10.1088/0004-637X/747/2/148}, \href
  {http://adsabs.harvard.edu/abs/2012ApJ...747..148D} {747, 148}

\bibitem[\protect\citeauthoryear{{Debes} et~al.,}{{Debes}
  et~al.}{2019}]{Debes2019}
{Debes} J.~H.,  et~al., 2019, \mn@doi [\apjl] {10.3847/2041-8213/ab0426}, \href
  {https://ui.adsabs.harvard.edu/abs/2019ApJ...872L..25D} {872, L25}

\bibitem[\protect\citeauthoryear{{Dennihy}, {Clemens}, {Dunlap}, {Fanale},
  {Fuchs}  \& {Hermes}}{{Dennihy} et~al.}{2018}]{Dennihy2018}
{Dennihy} E.,  {Clemens} J.~C.,  {Dunlap} B.~H.,  {Fanale} S.~M.,  {Fuchs}
  J.~T.,   {Hermes} J.~J.,  2018, \mn@doi [\apj] {10.3847/1538-4357/aaa89b},
  \href {http://adsabs.harvard.edu/abs/2018ApJ...854...40D} {854, 40}

\bibitem[\protect\citeauthoryear{{Dennihy} et~al.,}{{Dennihy}
  et~al.}{2020}]{Dennihy2020}
{Dennihy} E.,  et~al., 2020, \mn@doi [\apj] {10.3847/1538-4357/abc339}, \href
  {https://ui.adsabs.harvard.edu/abs/2020ApJ...905....5D} {905, 5}

\bibitem[\protect\citeauthoryear{{Dohnanyi}}{{Dohnanyi}}{1969}]{Dohnanyi1969}
{Dohnanyi} J.~S.,  1969, \mn@doi [\jgr] {10.1029/JB074i010p02531}, \href
  {http://adsabs.harvard.edu/abs/1969JGR....74.2531D} {74, 2531}

\bibitem[\protect\citeauthoryear{{Doyle}, {Young}, {Klein}, {Zuckerman}  \&
  {Schlichting}}{{Doyle} et~al.}{2019}]{Doyle2019}
{Doyle} A.~E.,  {Young} E.~D.,  {Klein} B.,  {Zuckerman} B.,   {Schlichting}
  H.~E.,  2019, \mn@doi [Science] {10.1126/science.aax3901}, \href
  {https://ui.adsabs.harvard.edu/abs/2019Sci...366..356D} {366, 356}

\bibitem[\protect\citeauthoryear{{Doyle}, {Klein}, {Schlichting}  \&
  {Young}}{{Doyle} et~al.}{2020}]{Doyle2020}
{Doyle} A.~E.,  {Klein} B.,  {Schlichting} H.~E.,   {Young} E.~D.,  2020,
  \mn@doi [\apj] {10.3847/1538-4357/abad9a}, \href
  {https://ui.adsabs.harvard.edu/abs/2020ApJ...901...10D} {901, 10}

\bibitem[\protect\citeauthoryear{{Duncan} \& {Lissauer}}{{Duncan} \&
  {Lissauer}}{1998}]{Duncan1998}
{Duncan} M.~J.,  {Lissauer} J.~J.,  1998, \mn@doi [\icarus]
  {10.1006/icar.1998.5962}, \href
  {http://adsabs.harvard.edu/abs/1998Icar..134..303D} {134, 303}

\bibitem[\protect\citeauthoryear{{Duvvuri}, {Redfield}  \& {Veras}}{{Duvvuri}
  et~al.}{2020}]{Duvvuri2020}
{Duvvuri} G.~M.,  {Redfield} S.,   {Veras} D.,  2020, \mn@doi [\apj]
  {10.3847/1538-4357/ab7fa0}, \href
  {https://ui.adsabs.harvard.edu/abs/2020ApJ...893..166D} {893, 166}

\bibitem[\protect\citeauthoryear{{Farihi}}{{Farihi}}{2016}]{Farihi2016}
{Farihi} J.,  2016, \mn@doi [\nar] {10.1016/j.newar.2016.03.001}, \href
  {http://adsabs.harvard.edu/abs/2016NewAR..71....9F} {71, 9}

\bibitem[\protect\citeauthoryear{{Farihi}, {Jura}  \& {Zuckerman}}{{Farihi}
  et~al.}{2009}]{Farihi2009}
{Farihi} J.,  {Jura} M.,   {Zuckerman} B.,  2009, \mn@doi [\apj]
  {10.1088/0004-637X/694/2/805}, \href
  {http://adsabs.harvard.edu/abs/2009ApJ...694..805F} {694, 805}

\bibitem[\protect\citeauthoryear{{Farihi}, {Barstow}, {Redfield}, {Dufour}  \&
  {Hambly}}{{Farihi} et~al.}{2010a}]{Farihi2010b}
{Farihi} J.,  {Barstow} M.~A.,  {Redfield} S.,  {Dufour} P.,   {Hambly} N.~C.,
  2010a, \mn@doi [\mnras] {10.1111/j.1365-2966.2010.16426.x}, \href
  {http://adsabs.harvard.edu/abs/2010MNRAS.404.2123F} {404, 2123}

\bibitem[\protect\citeauthoryear{{Farihi}, {Jura}, {Lee}  \&
  {Zuckerman}}{{Farihi} et~al.}{2010b}]{Farihi2010}
{Farihi} J.,  {Jura} M.,  {Lee} J.-E.,   {Zuckerman} B.,  2010b, \mn@doi [\apj]
  {10.1088/0004-637X/714/2/1386}, \href
  {http://adsabs.harvard.edu/abs/2010ApJ...714.1386F} {714, 1386}

\bibitem[\protect\citeauthoryear{{Farihi}, {G{\"a}nsicke}, {Wyatt}, {Girven},
  {Pringle}  \& {King}}{{Farihi} et~al.}{2012}]{Farihi2012}
{Farihi} J.,  {G{\"a}nsicke} B.~T.,  {Wyatt} M.~C.,  {Girven} J.,  {Pringle}
  J.~E.,   {King} A.~R.,  2012, \mn@doi [\mnras]
  {10.1111/j.1365-2966.2012.21215.x}, \href
  {http://adsabs.harvard.edu/abs/2012MNRAS.424..464F} {424, 464}

\bibitem[\protect\citeauthoryear{{Farihi}, {Parsons}  \&
  {G{\"a}nsicke}}{{Farihi} et~al.}{2017}]{Farihi2017}
{Farihi} J.,  {Parsons} S.~G.,   {G{\"a}nsicke} B.~T.,  2017, \mn@doi [Nature
  Astronomy] {10.1038/s41550-016-0032}, \href
  {http://adsabs.harvard.edu/abs/2017NatAs...1E..32F} {1, 0032}

\bibitem[\protect\citeauthoryear{{Fontaine} \& {Michaud}}{{Fontaine} \&
  {Michaud}}{1979}]{Fontaine1979}
{Fontaine} G.,  {Michaud} G.,  1979, \mn@doi [\apj] {10.1086/157247}, \href
  {http://adsabs.harvard.edu/abs/1979ApJ...231..826F} {231, 826}

\bibitem[\protect\citeauthoryear{{Foreman-Mackey}, {Hogg}  \&
  {Morton}}{{Foreman-Mackey} et~al.}{2014}]{ForemanMackey2014}
{Foreman-Mackey} D.,  {Hogg} D.~W.,   {Morton} T.~D.,  2014, \mn@doi [\apj]
  {10.1088/0004-637X/795/1/64}, \href
  {https://ui.adsabs.harvard.edu/abs/2014ApJ...795...64F} {795, 64}

\bibitem[\protect\citeauthoryear{{Fressin} et~al.,}{{Fressin}
  et~al.}{2013}]{Fressin2013}
{Fressin} F.,  et~al., 2013, \mn@doi [\apj] {10.1088/0004-637X/766/2/81}, \href
  {https://ui.adsabs.harvard.edu/abs/2013ApJ...766...81F} {766, 81}

\bibitem[\protect\citeauthoryear{{Frewen} \& {Hansen}}{{Frewen} \&
  {Hansen}}{2014}]{Frewen2014}
{Frewen} S.~F.~N.,  {Hansen} B.~M.~S.,  2014, \mn@doi [\mnras]
  {10.1093/mnras/stu097}, \href
  {http://adsabs.harvard.edu/abs/2014MNRAS.439.2442F} {439, 2442}

\bibitem[\protect\citeauthoryear{{Galicher} et~al.,}{{Galicher}
  et~al.}{2016}]{Galicher2016}
{Galicher} R.,  et~al., 2016, \mn@doi [\aap] {10.1051/0004-6361/201527828},
  \href {https://ui.adsabs.harvard.edu/abs/2016A&A...594A..63G} {594, A63}

\bibitem[\protect\citeauthoryear{{G{\"a}nsicke}, {Marsh}, {Southworth}  \&
  {Rebassa-Mansergas}}{{G{\"a}nsicke} et~al.}{2006}]{Gansicke2006}
{G{\"a}nsicke} B.~T.,  {Marsh} T.~R.,  {Southworth} J.,   {Rebassa-Mansergas}
  A.,  2006, \mn@doi [Science] {10.1126/science.1135033}, \href
  {http://adsabs.harvard.edu/abs/2006Sci...314.1908G} {314, 1908}

\bibitem[\protect\citeauthoryear{{G{\"a}nsicke} et~al.,}{{G{\"a}nsicke}
  et~al.}{2016}]{Gansicke2016}
{G{\"a}nsicke} B.~T.,  et~al., 2016, \mn@doi [\apjl]
  {10.3847/2041-8205/818/1/L7}, \href
  {http://adsabs.harvard.edu/abs/2016ApJ...818L...7G} {818, L7}

\bibitem[\protect\citeauthoryear{{Gary}, {Rappaport}, {Kaye}, {Alonso}  \&
  {Hambschs}}{{Gary} et~al.}{2017}]{Gary2017}
{Gary} B.~L.,  {Rappaport} S.,  {Kaye} T.~G.,  {Alonso} R.,   {Hambschs} F.-J.,
   2017, \mn@doi [\mnras] {10.1093/mnras/stw2921}, \href
  {http://adsabs.harvard.edu/abs/2017MNRAS.465.3267G} {465, 3267}

\bibitem[\protect\citeauthoryear{{Gentile Fusillo} et~al.,}{{Gentile Fusillo}
  et~al.}{2020}]{GentileFusillo2020}
{Gentile Fusillo} N.~P.,  et~al., 2020, arXiv e-prints, \href
  {https://ui.adsabs.harvard.edu/abs/2020arXiv201013807G} {p. arXiv:2010.13807}

\bibitem[\protect\citeauthoryear{{Gibson} et~al.,}{{Gibson}
  et~al.}{2009}]{Gibson2009}
{Gibson} N.~P.,  et~al., 2009, \mn@doi [\apj] {10.1088/0004-637X/700/2/1078},
  \href {https://ui.adsabs.harvard.edu/abs/2009ApJ...700.1078G} {700, 1078}

\bibitem[\protect\citeauthoryear{{Girven}, {Brinkworth}, {Farihi},
  {G{\"a}nsicke}, {Hoard}, {Marsh}  \& {Koester}}{{Girven}
  et~al.}{2012}]{Girven2012}
{Girven} J.,  {Brinkworth} C.~S.,  {Farihi} J.,  {G{\"a}nsicke} B.~T.,  {Hoard}
  D.~W.,  {Marsh} T.~R.,   {Koester} D.,  2012, \mn@doi [The Astrophysical
  Journal] {10.1088/0004-637X/749/2/154}, \href
  {https://ui.adsabs.harvard.edu/abs/2012ApJ...749..154G} {749, 154}

\bibitem[\protect\citeauthoryear{{Grishin} \& {Veras}}{{Grishin} \&
  {Veras}}{2019}]{Grishin2019}
{Grishin} E.,  {Veras} D.,  2019, \mn@doi [\mnras] {10.1093/mnras/stz2148},
  \href {https://ui.adsabs.harvard.edu/abs/2019MNRAS.tmp.2086G} {p.~2086}

\bibitem[\protect\citeauthoryear{{Guidry} et~al.,}{{Guidry}
  et~al.}{2020}]{Guidry2020}
{Guidry} J.~A.,  et~al., 2020, arXiv e-prints, \href
  {https://ui.adsabs.harvard.edu/abs/2020arXiv201200035G} {p. arXiv:2012.00035}

\bibitem[\protect\citeauthoryear{{Hamers} \& {Portegies Zwart}}{{Hamers} \&
  {Portegies Zwart}}{2016}]{Hamers2016}
{Hamers} A.~S.,  {Portegies Zwart} S.~F.,  2016, \mn@doi [\mnras]
  {10.1093/mnrasl/slw134}, \href
  {http://adsabs.harvard.edu/abs/2016MNRAS.462L..84H} {462, L84}

\bibitem[\protect\citeauthoryear{{Harrison}, {Bonsor}  \&
  {Madhusudhan}}{{Harrison} et~al.}{2018}]{Harrison2018}
{Harrison} J. H.~D.,  {Bonsor} A.,   {Madhusudhan} N.,  2018, \mn@doi [\mnras]
  {10.1093/mnras/sty1700}, \href
  {https://ui.adsabs.harvard.edu/abs/2018MNRAS.479.3814H} {479, 3814}

\bibitem[\protect\citeauthoryear{{Harrison}, {Shorttle}  \&
  {Bonsor}}{{Harrison} et~al.}{2021}]{Harrison2021}
{Harrison} J. H.~D.,  {Shorttle} O.,   {Bonsor} A.,  2021, \mn@doi [Earth and
  Planetary Science Letters] {10.1016/j.epsl.2020.116694}, \href
  {https://ui.adsabs.harvard.edu/abs/2021E&PSL.55416694H} {554, 116694}

\bibitem[\protect\citeauthoryear{{Hartman} et~al.,}{{Hartman}
  et~al.}{2012}]{Hartman2012}
{Hartman} J.~D.,  et~al., 2012, \mn@doi [\aj] {10.1088/0004-6256/144/5/139},
  \href {https://ui.adsabs.harvard.edu/abs/2012AJ....144..139H} {144, 139}

\bibitem[\protect\citeauthoryear{{Hellier} et~al.,}{{Hellier}
  et~al.}{2012}]{Hellier2012}
{Hellier} C.,  et~al., 2012, \mn@doi [\mnras]
  {10.1111/j.1365-2966.2012.21780.x}, \href
  {https://ui.adsabs.harvard.edu/abs/2012MNRAS.426..739H} {426, 739}

\bibitem[\protect\citeauthoryear{{Hollands}, {G{\"a}nsicke}  \&
  {Koester}}{{Hollands} et~al.}{2018}]{Hollands2018}
{Hollands} M.~A.,  {G{\"a}nsicke} B.~T.,   {Koester} D.,  2018, \mn@doi
  [\mnras] {10.1093/mnras/sty592}, \href
  {https://ui.adsabs.harvard.edu/abs/2018MNRAS.477...93H} {477, 93}

\bibitem[\protect\citeauthoryear{{Hsu}, {Ford}, {Ragozzine}  \&
  {Morehead}}{{Hsu} et~al.}{2018}]{Hsu2018}
{Hsu} D.~C.,  {Ford} E.~B.,  {Ragozzine} D.,   {Morehead} R.~C.,  2018, \mn@doi
  [\aj] {10.3847/1538-3881/aab9a8}, \href
  {https://ui.adsabs.harvard.edu/abs/2018AJ....155..205H} {155, 205}

\bibitem[\protect\citeauthoryear{{Ito} \& {Tanikawa}}{{Ito} \&
  {Tanikawa}}{1999}]{ItoTanikawa1999}
{Ito} T.,  {Tanikawa} K.,  1999, \mn@doi [\icarus] {10.1006/icar.1999.6112},
  \href {http://adsabs.harvard.edu/abs/1999Icar..139..336I} {139, 336}

\bibitem[\protect\citeauthoryear{{Johnson}, {Aller}, {Howard}  \&
  {Crepp}}{{Johnson} et~al.}{2010}]{Johnson2010}
{Johnson} J.~A.,  {Aller} K.~M.,  {Howard} A.~W.,   {Crepp} J.~R.,  2010,
  \mn@doi [\pasp] {10.1086/655775}, \href
  {https://ui.adsabs.harvard.edu/abs/2010PASP..122..905J} {122, 905}

\bibitem[\protect\citeauthoryear{{Jura}}{{Jura}}{2003}]{Jura2003}
{Jura} M.,  2003, \mn@doi [\apjl] {10.1086/374036}, \href
  {http://adsabs.harvard.edu/abs/2003ApJ...584L..91J} {584, L91}

\bibitem[\protect\citeauthoryear{{Jura}}{{Jura}}{2006}]{Jura2006}
{Jura} M.,  2006, \mn@doi [\apj] {10.1086/508738}, \href
  {http://adsabs.harvard.edu/abs/2006ApJ...653..613J} {653, 613}

\bibitem[\protect\citeauthoryear{{Jura}, {Farihi}  \& {Zuckerman}}{{Jura}
  et~al.}{2007}]{Jura2007}
{Jura} M.,  {Farihi} J.,   {Zuckerman} B.,  2007, \mn@doi [\apj]
  {10.1086/518767}, \href
  {https://ui.adsabs.harvard.edu/abs/2007ApJ...663.1285J} {663, 1285}

\bibitem[\protect\citeauthoryear{{Jura}, {Muno}, {Farihi}  \&
  {Zuckerman}}{{Jura} et~al.}{2009}]{Jura2009}
{Jura} M.,  {Muno} M.~P.,  {Farihi} J.,   {Zuckerman} B.,  2009, \mn@doi [\apj]
  {10.1088/0004-637X/699/2/1473}, \href
  {http://adsabs.harvard.edu/abs/2009ApJ...699.1473J} {699, 1473}

\bibitem[\protect\citeauthoryear{{Kaiser}, {Clemens}, {Blouin}, {Dufour},
  {Hegedus}, {Reding}  \& {B{\'e}dard}}{{Kaiser} et~al.}{2021}]{Kaiser2021}
{Kaiser} B.~C.,  {Clemens} J.~C.,  {Blouin} S.,  {Dufour} P.,  {Hegedus} R.~J.,
   {Reding} J.~S.,   {B{\'e}dard} A.,  2021, \mn@doi [Science]
  {10.1126/science.abd1714}, \href
  {https://ui.adsabs.harvard.edu/abs/2021Sci...371..168K} {371, 168}

\bibitem[\protect\citeauthoryear{{Kalas} et~al.,}{{Kalas}
  et~al.}{2008}]{Kalas2008}
{Kalas} P.,  et~al., 2008, \mn@doi [Science] {10.1126/science.1166609}, \href
  {https://ui.adsabs.harvard.edu/abs/2008Sci...322.1345K} {322, 1345}

\bibitem[\protect\citeauthoryear{{Kilic} \& {Redfield}}{{Kilic} \&
  {Redfield}}{2007}]{Kilic2007}
{Kilic} M.,  {Redfield} S.,  2007, \mn@doi [\apj] {10.1086/513008}, \href
  {http://adsabs.harvard.edu/abs/2007ApJ...660..641K} {660, 641}

\bibitem[\protect\citeauthoryear{{Kilic}, {von Hippel}, {Leggett}  \&
  {Winget}}{{Kilic} et~al.}{2006}]{Kilic2006}
{Kilic} M.,  {von Hippel} T.,  {Leggett} S.~K.,   {Winget} D.~E.,  2006,
  \mn@doi [\apj] {10.1086/504682}, \href
  {http://adsabs.harvard.edu/abs/2006ApJ...646..474K} {646, 474}

\bibitem[\protect\citeauthoryear{{Kipping} \& {Sandford}}{{Kipping} \&
  {Sandford}}{2016}]{Kipping2016}
{Kipping} D.~M.,  {Sandford} E.,  2016, \mn@doi [\mnras]
  {10.1093/mnras/stw1926}, \href
  {https://ui.adsabs.harvard.edu/abs/2016MNRAS.463.1323K} {463, 1323}

\bibitem[\protect\citeauthoryear{{Klein}, {Jura}, {Koester}, {Zuckerman}  \&
  {Melis}}{{Klein} et~al.}{2010}]{Klein2010}
{Klein} B.,  {Jura} M.,  {Koester} D.,  {Zuckerman} B.,   {Melis} C.,  2010,
  \mn@doi [\apj] {10.1088/0004-637X/709/2/950}, \href
  {http://adsabs.harvard.edu/abs/2010ApJ...709..950K} {709, 950}

\bibitem[\protect\citeauthoryear{{Klein}, {Doyle}, {Zuckerman}, {Dufour},
  {Blouin}, {Melis}, {Weinberger}  \& {Young}}{{Klein}
  et~al.}{2021}]{Klein2021}
{Klein} B.,  {Doyle} A.~E.,  {Zuckerman} B.,  {Dufour} P.,  {Blouin} S.,
  {Melis} C.,  {Weinberger} A.~J.,   {Young} E.~D.,  2021, arXiv e-prints,
  \href {https://ui.adsabs.harvard.edu/abs/2021arXiv210201834K} {p.
  arXiv:2102.01834}

\bibitem[\protect\citeauthoryear{{Koester}}{{Koester}}{2009}]{Koester2009}
{Koester} D.,  2009, \mn@doi [\aap] {10.1051/0004-6361/200811468}, \href
  {http://adsabs.harvard.edu/abs/2009A%26A...498..517K} {498, 517}

\bibitem[\protect\citeauthoryear{{Koester} \& {Wilken}}{{Koester} \&
  {Wilken}}{2006}]{Koester2006}
{Koester} D.,  {Wilken} D.,  2006, \mn@doi [\aap] {10.1051/0004-6361:20064843},
  \href {http://adsabs.harvard.edu/abs/2006A%26A...453.1051K} {453, 1051}

\bibitem[\protect\citeauthoryear{{Koester}, {G{\"a}nsicke}  \&
  {Farihi}}{{Koester} et~al.}{2014}]{Koester2014}
{Koester} D.,  {G{\"a}nsicke} B.~T.,   {Farihi} J.,  2014, \mn@doi [\aap]
  {10.1051/0004-6361/201423691}, \href
  {http://adsabs.harvard.edu/abs/2014A%26A...566A..34K} {566, A34}

\bibitem[\protect\citeauthoryear{{Kratter} \& {Perets}}{{Kratter} \&
  {Perets}}{2012}]{Kratter2012}
{Kratter} K.~M.,  {Perets} H.~B.,  2012, \mn@doi [\apj]
  {10.1088/0004-637X/753/1/91}, \href
  {https://ui.adsabs.harvard.edu/abs/2012ApJ...753...91K} {753, 91}

\bibitem[\protect\citeauthoryear{{Lagrange} et~al.,}{{Lagrange}
  et~al.}{2010}]{Lagrange2010}
{Lagrange} A.~M.,  et~al., 2010, \mn@doi [Science] {10.1126/science.1187187},
  \href {https://ui.adsabs.harvard.edu/abs/2010Sci...329...57L} {329, 57}

\bibitem[\protect\citeauthoryear{{Latham} et~al.,}{{Latham}
  et~al.}{2011}]{Latham2011}
{Latham} D.~W.,  et~al., 2011, \mn@doi [\apjl] {10.1088/2041-8205/732/2/L24},
  \href {https://ui.adsabs.harvard.edu/abs/2011ApJ...732L..24L} {732, L24}

\bibitem[\protect\citeauthoryear{{Lecar}, {Podolak}, {Sasselov}  \&
  {Chiang}}{{Lecar} et~al.}{2006}]{Lecar2006}
{Lecar} M.,  {Podolak} M.,  {Sasselov} D.,   {Chiang} E.,  2006, \mn@doi [\apj]
  {10.1086/500287}, \href {http://adsabs.harvard.edu/abs/2006ApJ...640.1115L}
  {640, 1115}

\bibitem[\protect\citeauthoryear{{Makarov} \& {Veras}}{{Makarov} \&
  {Veras}}{2019}]{Valeri2019}
{Makarov} V.~V.,  {Veras} D.,  2019, arXiv e-prints, \href
  {https://ui.adsabs.harvard.edu/abs/2019arXiv190804612M} {p. arXiv:1908.04612}

\bibitem[\protect\citeauthoryear{{Malavolta} et~al.,}{{Malavolta}
  et~al.}{2016}]{Malavolta2016}
{Malavolta} L.,  et~al., 2016, \mn@doi [\aap] {10.1051/0004-6361/201527933},
  \href {https://ui.adsabs.harvard.edu/abs/2016A&A...588A.118M} {588, A118}

\bibitem[\protect\citeauthoryear{{Malhotra}}{{Malhotra}}{1999}]{Malhotra1999}
{Malhotra} R.,  1999, \mn@doi [\nat] {10.1038/45123}, \href
  {http://adsabs.harvard.edu/abs/1999Natur.402..599M} {402, 599}

\bibitem[\protect\citeauthoryear{Malhotra}{Malhotra}{2012}]{malhotra2012}
Malhotra R.,  2012, Encyclopedia of Life Support Systems by UNESCO, 6, 55

\bibitem[\protect\citeauthoryear{{Malhotra}, {Fox}, {Murray}  \&
  {Nicholson}}{{Malhotra} et~al.}{1989}]{Malhotra1989}
{Malhotra} R.,  {Fox} K.,  {Murray} C.~D.,   {Nicholson} P.~D.,  1989, \aap,
  \href {http://adsabs.harvard.edu/abs/1989A%26A...221..348M} {221, 348}

\bibitem[\protect\citeauthoryear{{Manser} et~al.,}{{Manser}
  et~al.}{2019}]{Manser2019}
{Manser} C.~J.,  et~al., 2019, \mn@doi [Science] {10.1126/science.aat5330},
  \href {https://ui.adsabs.harvard.edu/abs/2019Sci...364...66M} {364, 66}

\bibitem[\protect\citeauthoryear{{Manser}, {G{\"a}nsicke}, {Gentile Fusillo},
  {Ashley}, {Breedt}, {Hollands}, {Izquierdo}  \& {Pelisoli}}{{Manser}
  et~al.}{2020}]{Manser2020}
{Manser} C.~J.,  {G{\"a}nsicke} B.~T.,  {Gentile Fusillo} N.~P.,  {Ashley} R.,
  {Breedt} E.,  {Hollands} M.,  {Izquierdo} P.,   {Pelisoli} I.,  2020, \mn@doi
  [\mnras] {10.1093/mnras/staa359}, \href
  {https://ui.adsabs.harvard.edu/abs/2020MNRAS.493.2127M} {493, 2127}

\bibitem[\protect\citeauthoryear{{Marois}, {Macintosh}, {Barman}, {Zuckerman},
  {Song}, {Patience}, {Lafreni{\`e}re}  \& {Doyon}}{{Marois}
  et~al.}{2008}]{Marois2008}
{Marois} C.,  {Macintosh} B.,  {Barman} T.,  {Zuckerman} B.,  {Song} I.,
  {Patience} J.,  {Lafreni{\`e}re} D.,   {Doyon} R.,  2008, \mn@doi [Science]
  {10.1126/science.1166585}, \href
  {https://ui.adsabs.harvard.edu/abs/2008Sci...322.1348M} {322, 1348}

\bibitem[\protect\citeauthoryear{{Marois}, {Zuckerman}, {Konopacky},
  {Macintosh}  \& {Barman}}{{Marois} et~al.}{2010}]{Marois2010}
{Marois} C.,  {Zuckerman} B.,  {Konopacky} Q.~M.,  {Macintosh} B.,   {Barman}
  T.,  2010, \mn@doi [\nat] {10.1038/nature09684}, \href
  {http://adsabs.harvard.edu/abs/2010Natur.468.1080M} {468, 1080}

\bibitem[\protect\citeauthoryear{{Martin} \& {Livio}}{{Martin} \&
  {Livio}}{2012}]{Martin2012snowline}
{Martin} R.~G.,  {Livio} M.,  2012, \mn@doi [\mnras]
  {10.1111/j.1745-3933.2012.01290.x}, \href
  {https://ui.adsabs.harvard.edu/abs/2012MNRAS.425L...6M} {425, L6}

\bibitem[\protect\citeauthoryear{{Martin} \& {Livio}}{{Martin} \&
  {Livio}}{2013a}]{Martin2013asteroids}
{Martin} R.~G.,  {Livio} M.,  2013a, \mn@doi [\mnras] {10.1093/mnrasl/sls003},
  \href {https://ui.adsabs.harvard.edu/abs/2013MNRAS.428L..11M} {428, L11}

\bibitem[\protect\citeauthoryear{{Martin} \& {Livio}}{{Martin} \&
  {Livio}}{2013b}]{Martin2013}
{Martin} R.~G.,  {Livio} M.,  2013b, \mn@doi [\mnras] {10.1093/mnras/stt1051},
  \href {http://adsabs.harvard.edu/abs/2013MNRAS.434..633M} {434, 633}

\bibitem[\protect\citeauthoryear{{Martin} \& {Livio}}{{Martin} \&
  {Livio}}{2015}]{Martin2015}
{Martin} R.~G.,  {Livio} M.,  2015, \mn@doi [\apj]
  {10.1088/0004-637X/810/2/105}, \href
  {https://ui.adsabs.harvard.edu/abs/2015ApJ...810..105M} {810, 105}

\bibitem[\protect\citeauthoryear{{Martin} \& {Livio}}{{Martin} \&
  {Livio}}{2016}]{MartinLivio2016}
{Martin} R.~G.,  {Livio} M.,  2016, \mn@doi [\apj]
  {10.3847/0004-637X/822/2/90}, \href
  {http://adsabs.harvard.edu/abs/2016ApJ...822...90M} {822, 90}

\bibitem[\protect\citeauthoryear{{Martin}, {Livio}, {Smallwood}  \&
  {Chen}}{{Martin} et~al.}{2020}]{Martin2020}
{Martin} R.~G.,  {Livio} M.,  {Smallwood} J.~L.,   {Chen} C.,  2020, \mn@doi
  [\mnras] {10.1093/mnrasl/slaa030}, \href
  {https://ui.adsabs.harvard.edu/abs/2020MNRAS.494L..17M} {494, L17}

\bibitem[\protect\citeauthoryear{{Mayor} et~al.,}{{Mayor}
  et~al.}{2011}]{Mayor2011}
{Mayor} M.,  et~al., 2011, arXiv e-prints, \href
  {https://ui.adsabs.harvard.edu/abs/2011arXiv1109.2497M} {p. arXiv:1109.2497}

\bibitem[\protect\citeauthoryear{{McDonald} \& {Zijlstra}}{{McDonald} \&
  {Zijlstra}}{2015}]{McDonald2015}
{McDonald} I.,  {Zijlstra} A.~A.,  2015, \mn@doi [\mnras]
  {10.1093/mnras/stv007}, \href
  {http://adsabs.harvard.edu/abs/2015MNRAS.448..502M} {448, 502}

\bibitem[\protect\citeauthoryear{{Melis} \& {Dufour}}{{Melis} \&
  {Dufour}}{2017}]{Melis2017}
{Melis} C.,  {Dufour} P.,  2017, \mn@doi [\apj] {10.3847/1538-4357/834/1/1},
  \href {http://adsabs.harvard.edu/abs/2017ApJ...834....1M} {834, 1}

\bibitem[\protect\citeauthoryear{{Melis}, {Jura}, {Albert}, {Klein}  \&
  {Zuckerman}}{{Melis} et~al.}{2010}]{Melis2010}
{Melis} C.,  {Jura} M.,  {Albert} L.,  {Klein} B.,   {Zuckerman} B.,  2010,
  \mn@doi [\apj] {10.1088/0004-637X/722/2/1078}, \href
  {http://adsabs.harvard.edu/abs/2010ApJ...722.1078M} {722, 1078}

\bibitem[\protect\citeauthoryear{{Melis}, {Klein}, {Doyle}, {Weinberger},
  {Zuckerman}  \& {Dufour}}{{Melis} et~al.}{2020}]{Melis2020}
{Melis} C.,  {Klein} B.,  {Doyle} A.~E.,  {Weinberger} A.,  {Zuckerman} B.,
  {Dufour} P.,  2020, \mn@doi [\apj] {10.3847/1538-4357/abbdfa}, \href
  {https://ui.adsabs.harvard.edu/abs/2020ApJ...905...56M} {905, 56}

\bibitem[\protect\citeauthoryear{{Ment}, {Fischer}, {Bakos}, {Howard}  \&
  {Isaacson}}{{Ment} et~al.}{2018}]{Ment2018}
{Ment} K.,  {Fischer} D.~A.,  {Bakos} G.,  {Howard} A.~W.,   {Isaacson} H.,
  2018, \mn@doi [\aj] {10.3847/1538-3881/aae1f5}, \href
  {https://ui.adsabs.harvard.edu/abs/2018AJ....156..213M} {156, 213}

\bibitem[\protect\citeauthoryear{{Meshkat} et~al.,}{{Meshkat}
  et~al.}{2017}]{Meshkat2017}
{Meshkat} T.,  et~al., 2017, \mn@doi [\aj] {10.3847/1538-3881/aa8e9a}, \href
  {https://ui.adsabs.harvard.edu/abs/2017AJ....154..245M} {154, 245}

\bibitem[\protect\citeauthoryear{{Milani} \& {Knezevic}}{{Milani} \&
  {Knezevic}}{1990}]{Milani1990}
{Milani} A.,  {Knezevic} Z.,  1990, \mn@doi [Celestial Mechanics and Dynamical
  Astronomy] {10.1007/BF00049444}, \href
  {http://adsabs.harvard.edu/abs/1990CeMDA..49..347M} {49, 347}

\bibitem[\protect\citeauthoryear{{Moons} \& {Morbidelli}}{{Moons} \&
  {Morbidelli}}{1995}]{moons1995}
{Moons} M.,  {Morbidelli} A.,  1995, \mn@doi [\icarus]
  {10.1006/icar.1995.1041}, \href
  {http://adsabs.harvard.edu/abs/1995Icar..114...33M} {114, 33}

\bibitem[\protect\citeauthoryear{{Morales}, {Rieke}, {Werner}, {Bryden},
  {Stapelfeldt}  \& {Su}}{{Morales} et~al.}{2011}]{Morales2011}
{Morales} F.~Y.,  {Rieke} G.~H.,  {Werner} M.~W.,  {Bryden} G.,  {Stapelfeldt}
  K.~R.,   {Su} K.~Y.~L.,  2011, \mn@doi [\apjl] {10.1088/2041-8205/730/2/L29},
  \href {http://adsabs.harvard.edu/abs/2011ApJ...730L..29M} {730, L29}

\bibitem[\protect\citeauthoryear{{Morbidelli} \& {Henrard}}{{Morbidelli} \&
  {Henrard}}{1991}]{morbidelli1991}
{Morbidelli} A.,  {Henrard} J.,  1991, \mn@doi [Celestial Mechanics and
  Dynamical Astronomy] {10.1007/BF00048607}, \href
  {http://adsabs.harvard.edu/abs/1991CeMDA..51..169M} {51, 169}

\bibitem[\protect\citeauthoryear{{Morton} \& {Johnson}}{{Morton} \&
  {Johnson}}{2011}]{Morton2011}
{Morton} T.~D.,  {Johnson} J.~A.,  2011, \mn@doi [\apj]
  {10.1088/0004-637X/729/2/138}, \href
  {https://ui.adsabs.harvard.edu/abs/2011ApJ...729..138M} {729, 138}

\bibitem[\protect\citeauthoryear{{Mustill} \& {Villaver}}{{Mustill} \&
  {Villaver}}{2012}]{Mustill2012}
{Mustill} A.~J.,  {Villaver} E.,  2012, \mn@doi [\apj]
  {10.1088/0004-637X/761/2/121}, \href
  {http://adsabs.harvard.edu/abs/2012ApJ...761..121M} {761, 121}

\bibitem[\protect\citeauthoryear{{Mustill}, {Villaver}, {Veras}, {G{\"a}nsicke}
   \& {Bonsor}}{{Mustill} et~al.}{2018}]{Mustill2018}
{Mustill} A.~J.,  {Villaver} E.,  {Veras} D.,  {G{\"a}nsicke} B.~T.,   {Bonsor}
  A.,  2018, \mn@doi [Monthly Notices of the Royal Astronomical Society]
  {10.1093/mnras/sty446}, \href
  {https://ui.adsabs.harvard.edu/abs/2018MNRAS.476.3939M} {476, 3939}

\bibitem[\protect\citeauthoryear{{Neveu-VanMalle} et~al.,}{{Neveu-VanMalle}
  et~al.}{2016}]{NeveuVanMalle2016}
{Neveu-VanMalle} M.,  et~al., 2016, \mn@doi [\aap]
  {10.1051/0004-6361/201526965}, \href
  {https://ui.adsabs.harvard.edu/abs/2016A&A...586A..93N} {586, A93}

\bibitem[\protect\citeauthoryear{{Nielsen} et~al.,}{{Nielsen}
  et~al.}{2019}]{Nielsen2019}
{Nielsen} L.~D.,  et~al., 2019, \mn@doi [\mnras] {10.1093/mnras/stz2351}, \href
  {https://ui.adsabs.harvard.edu/abs/2019MNRAS.489.2478N} {489, 2478}

\bibitem[\protect\citeauthoryear{{O'Connor} \& {Lai}}{{O'Connor} \&
  {Lai}}{2020}]{OConnor2020}
{O'Connor} C.~E.,  {Lai} D.,  2020, \mn@doi [\mnras] {10.1093/mnras/staa2645},
  \href {https://ui.adsabs.harvard.edu/abs/2020MNRAS.498.4005O} {498, 4005}

\bibitem[\protect\citeauthoryear{{Paquette}, {Pelletier}, {Fontaine}  \&
  {Michaud}}{{Paquette} et~al.}{1986}]{Paquette1986}
{Paquette} C.,  {Pelletier} C.,  {Fontaine} G.,   {Michaud} G.,  1986, \mn@doi
  [\apjs] {10.1086/191112}, \href
  {http://adsabs.harvard.edu/abs/1986ApJS...61..197P} {61, 197}

\bibitem[\protect\citeauthoryear{{Payne}, {Veras}, {Holman}  \&
  {G{\"a}nsicke}}{{Payne} et~al.}{2016}]{Payne2016}
{Payne} M.~J.,  {Veras} D.,  {Holman} M.~J.,   {G{\"a}nsicke} B.~T.,  2016,
  \mn@doi [\mnras] {10.1093/mnras/stv2966}, \href
  {http://adsabs.harvard.edu/abs/2016MNRAS.457..217P} {457, 217}

\bibitem[\protect\citeauthoryear{{Payne}, {Veras}, {G{\"a}nsicke}  \&
  {Holman}}{{Payne} et~al.}{2017}]{Payne2017}
{Payne} M.~J.,  {Veras} D.,  {G{\"a}nsicke} B.~T.,   {Holman} M.~J.,  2017,
  \mn@doi [\mnras] {10.1093/mnras/stw2585}, \href
  {http://adsabs.harvard.edu/abs/2017MNRAS.464.2557P} {464, 2557}

\bibitem[\protect\citeauthoryear{{Petigura}, {Howard}  \& {Marcy}}{{Petigura}
  et~al.}{2013}]{Petigura2013}
{Petigura} E.~A.,  {Howard} A.~W.,   {Marcy} G.~W.,  2013, \mn@doi [Proceedings
  of the National Academy of Science] {10.1073/pnas.1319909110}, \href
  {https://ui.adsabs.harvard.edu/abs/2013PNAS..11019273P} {110, 19273}

\bibitem[\protect\citeauthoryear{{Petrovich} \& {Mu{\~n}oz}}{{Petrovich} \&
  {Mu{\~n}oz}}{2017}]{Petrovich2017}
{Petrovich} C.,  {Mu{\~n}oz} D.~J.,  2017, \mn@doi [\apj]
  {10.3847/1538-4357/834/2/116}, \href
  {http://adsabs.harvard.edu/abs/2017ApJ...834..116P} {834, 116}

\bibitem[\protect\citeauthoryear{{Pollack}, {Hubickyj}, {Bodenheimer},
  {Lissauer}, {Podolak}  \& {Greenzweig}}{{Pollack} et~al.}{1996}]{Pollack1996}
{Pollack} J.~B.,  {Hubickyj} O.,  {Bodenheimer} P.,  {Lissauer} J.~J.,
  {Podolak} M.,   {Greenzweig} Y.,  1996, \mn@doi [\icarus]
  {10.1006/icar.1996.0190}, \href
  {https://ui.adsabs.harvard.edu/abs/1996Icar..124...62P} {124, 62}

\bibitem[\protect\citeauthoryear{{Rappaport}, {Gary}, {Kaye}, {Vanderburg},
  {Croll}, {Benni}  \& {Foote}}{{Rappaport} et~al.}{2016}]{Rappaport2016}
{Rappaport} S.,  {Gary} B.~L.,  {Kaye} T.,  {Vanderburg} A.,  {Croll} B.,
  {Benni} P.,   {Foote} J.,  2016, \mn@doi [\mnras] {10.1093/mnras/stw612},
  \href {http://adsabs.harvard.edu/abs/2016MNRAS.458.3904R} {458, 3904}

\bibitem[\protect\citeauthoryear{{Rebassa-Mansergas}, {Solano}, {Xu},
  {Rodrigo}, {Jim{\'e}nez-Esteban}  \& {Torres}}{{Rebassa-Mansergas}
  et~al.}{2019}]{RebassaMansergas2019}
{Rebassa-Mansergas} A.,  {Solano} E.,  {Xu} S.,  {Rodrigo} C.,
  {Jim{\'e}nez-Esteban} F.~M.,   {Torres} S.,  2019, \mn@doi [\mnras]
  {10.1093/mnras/stz2423}, \href
  {https://ui.adsabs.harvard.edu/abs/2019MNRAS.489.3990R} {489, 3990}

\bibitem[\protect\citeauthoryear{{Redfield}, {Farihi}, {Cauley}, {Parsons},
  {G{\"a}nsicke}  \& {Duvvuri}}{{Redfield} et~al.}{2017}]{Redfield2017}
{Redfield} S.,  {Farihi} J.,  {Cauley} P.~W.,  {Parsons} S.~G.,  {G{\"a}nsicke}
  B.~T.,   {Duvvuri} G.~M.,  2017, \mn@doi [\apj] {10.3847/1538-4357/aa68a0},
  \href {http://adsabs.harvard.edu/abs/2017ApJ...839...42R} {839, 42}

\bibitem[\protect\citeauthoryear{{Reffert}, {Bergmann}, {Quirrenbach},
  {Trifonov}  \& {Kuenstler}}{{Reffert} et~al.}{2014}]{Reffert2014}
{Reffert} S.,  {Bergmann} C.,  {Quirrenbach} A.,  {Trifonov} T.,   {Kuenstler}
  A.,  2014, VizieR Online Data Catalog, \href
  {https://ui.adsabs.harvard.edu/abs/2014yCat..35740116R} {pp J/A+A/574/A116}

\bibitem[\protect\citeauthoryear{{Reffert}, {Bergmann}, {Quirrenbach},
  {Trifonov}  \& {K{\"u}nstler}}{{Reffert} et~al.}{2015}]{Reffert2015}
{Reffert} S.,  {Bergmann} C.,  {Quirrenbach} A.,  {Trifonov} T.,
  {K{\"u}nstler} A.,  2015, \mn@doi [\aap] {10.1051/0004-6361/201322360}, \href
  {https://ui.adsabs.harvard.edu/abs/2015A&A...574A.116R} {574, A116}

\bibitem[\protect\citeauthoryear{{Reimers}}{{Reimers}}{1977}]{Reimers1977}
{Reimers} D.,  1977, \aap, \href
  {http://adsabs.harvard.edu/abs/1977A%26A....54..485R} {54, 485}

\bibitem[\protect\citeauthoryear{{Ronco}, {Schreiber}, {Giuppone}, {Veras},
  {Cuadra}  \& {Guilera}}{{Ronco} et~al.}{2020}]{Ronco2020}
{Ronco} M.~P.,  {Schreiber} M.~R.,  {Giuppone} C.~A.,  {Veras} D.,  {Cuadra}
  J.,   {Guilera} O.~M.,  2020, \mn@doi [\apjl] {10.3847/2041-8213/aba35f},
  \href {https://ui.adsabs.harvard.edu/abs/2020ApJ...898L..23R} {898, L23}

\bibitem[\protect\citeauthoryear{{Rosenfield} et~al.,}{{Rosenfield}
  et~al.}{2014}]{Rosenfield2014}
{Rosenfield} P.,  et~al., 2014, \mn@doi [\apj] {10.1088/0004-637X/790/1/22},
  \href {http://adsabs.harvard.edu/abs/2014ApJ...790...22R} {790, 22}

\bibitem[\protect\citeauthoryear{{Rosenfield}, {Marigo}, {Girardi},
  {Dalcanton}, {Bressan}, {Williams}  \& {Dolphin}}{{Rosenfield}
  et~al.}{2016}]{Rosenfield2016}
{Rosenfield} P.,  {Marigo} P.,  {Girardi} L.,  {Dalcanton} J.~J.,  {Bressan}
  A.,  {Williams} B.~F.,   {Dolphin} A.,  2016, \mn@doi [\apj]
  {10.3847/0004-637X/822/2/73}, \href
  {http://adsabs.harvard.edu/abs/2016ApJ...822...73R} {822, 73}

\bibitem[\protect\citeauthoryear{{Schr{\"o}der} \& {Connon
  Smith}}{{Schr{\"o}der} \& {Connon Smith}}{2008}]{Schroder2008}
{Schr{\"o}der} K.-P.,  {Connon Smith} R.,  2008, \mn@doi [\mnras]
  {10.1111/j.1365-2966.2008.13022.x}, \href
  {http://adsabs.harvard.edu/abs/2008MNRAS.386..155S} {386, 155}

\bibitem[\protect\citeauthoryear{{Siess} \& {Livio}}{{Siess} \&
  {Livio}}{1999}]{Siess1999}
{Siess} L.,  {Livio} M.,  1999, \mn@doi [\mnras]
  {10.1046/j.1365-8711.1999.02784.x}, \href
  {http://adsabs.harvard.edu/abs/1999MNRAS.308.1133S} {308, 1133}

\bibitem[\protect\citeauthoryear{{Smallwood}, {Martin}, {Lepp}  \&
  {Livio}}{{Smallwood} et~al.}{2018a}]{smallwood2018a}
{Smallwood} J.~L.,  {Martin} R.~G.,  {Lepp} S.,   {Livio} M.,  2018a, \mn@doi
  [\mnras] {10.1093/mnras/stx2384}, \href
  {http://adsabs.harvard.edu/abs/2018MNRAS.473..295S} {473, 295}

\bibitem[\protect\citeauthoryear{{Smallwood}, {Martin}, {Livio}  \&
  {Lubow}}{{Smallwood} et~al.}{2018b}]{smallwood2018b}
{Smallwood} J.~L.,  {Martin} R.~G.,  {Livio} M.,   {Lubow} S.~H.,  2018b,
  \mn@doi [\mnras] {10.1093/mnras/sty1819}, \href
  {http://adsabs.harvard.edu/abs/2018MNRAS.480...57S} {480, 57}

\bibitem[\protect\citeauthoryear{{Stassun}, {Collins}  \& {Gaudi}}{{Stassun}
  et~al.}{2017}]{Stassum2017}
{Stassun} K.~G.,  {Collins} K.~A.,   {Gaudi} B.~S.,  2017, \mn@doi [\aj]
  {10.3847/1538-3881/aa5df3}, \href
  {https://ui.adsabs.harvard.edu/abs/2017AJ....153..136S} {153, 136}

\bibitem[\protect\citeauthoryear{{Steffen} et~al.,}{{Steffen}
  et~al.}{2012}]{Steffen2012}
{Steffen} J.~H.,  et~al., 2012, \mn@doi [Proceedings of the National Academy of
  Science] {10.1073/pnas.1120970109}, \href
  {https://ui.adsabs.harvard.edu/abs/2012PNAS..109.7982S} {109, 7982}

\bibitem[\protect\citeauthoryear{{Stephan}, {Naoz}  \& {Zuckerman}}{{Stephan}
  et~al.}{2017}]{Stephan2017}
{Stephan} A.~P.,  {Naoz} S.,   {Zuckerman} B.,  2017, \mn@doi [The
  Astrophysical Journal] {10.3847/2041-8213/aa7cf3}, \href
  {https://ui.adsabs.harvard.edu/abs/2017ApJ...844L..16S} {844, L16}

\bibitem[\protect\citeauthoryear{{Stephan}, {Naoz}  \& {Gaudi}}{{Stephan}
  et~al.}{2018}]{Stephan2018}
{Stephan} A.~P.,  {Naoz} S.,   {Gaudi} B.~S.,  2018, \mn@doi [The Astronomical
  Journal] {10.3847/1538-3881/aad6e5}, \href
  {https://ui.adsabs.harvard.edu/abs/2018AJ....156..128S} {156, 128}

\bibitem[\protect\citeauthoryear{{Stone}, {Metzger}  \& {Loeb}}{{Stone}
  et~al.}{2015}]{Stone2015}
{Stone} N.,  {Metzger} B.~D.,   {Loeb} A.,  2015, \mn@doi [Monthly Notices of
  the Royal Astronomical Society] {10.1093/mnras/stu2718}, \href
  {https://ui.adsabs.harvard.edu/abs/2015MNRAS.448..188S} {448, 188}

\bibitem[\protect\citeauthoryear{{Swan}, {Farihi}, {Koester}, {Hollands},
  {Parsons}, {Cauley}, {Redfield}  \& {G{\"a}nsicke}}{{Swan}
  et~al.}{2019}]{Swan2019}
{Swan} A.,  {Farihi} J.,  {Koester} D.,  {Hollands} M.,  {Parsons} S.,
  {Cauley} P.~W.,  {Redfield} S.,   {G{\"a}nsicke} B.~T.,  2019, \mn@doi
  [\mnras] {10.1093/mnras/stz2337}, \href
  {https://ui.adsabs.harvard.edu/abs/2019MNRAS.490..202S} {490, 202}

\bibitem[\protect\citeauthoryear{{Tremblay}, {Cummings}, {Kalirai},
  {G{\"a}nsicke}, {Gentile-Fusillo}  \& {Raddi}}{{Tremblay}
  et~al.}{2016}]{Tremblay2016}
{Tremblay} P.-E.,  {Cummings} J.,  {Kalirai} J.~S.,  {G{\"a}nsicke} B.~T.,
  {Gentile-Fusillo} N.,   {Raddi} R.,  2016, \mn@doi [\mnras]
  {10.1093/mnras/stw1447}, \href
  {http://adsabs.harvard.edu/abs/2016MNRAS.461.2100T} {461, 2100}

\bibitem[\protect\citeauthoryear{{Triaud} et~al.,}{{Triaud}
  et~al.}{2010}]{Triaud2010}
{Triaud} A.~H.~M.~J.,  et~al., 2010, \mn@doi [\aap]
  {10.1051/0004-6361/201014525}, \href
  {https://ui.adsabs.harvard.edu/abs/2010A&A...524A..25T} {524, A25}

\bibitem[\protect\citeauthoryear{{Vanderbosch} et~al.,}{{Vanderbosch}
  et~al.}{2020}]{Vanderbosch2020}
{Vanderbosch} Z.,  et~al., 2020, \mn@doi [\apj] {10.3847/1538-4357/ab9649},
  \href {https://ui.adsabs.harvard.edu/abs/2020ApJ...897..171V} {897, 171}

\bibitem[\protect\citeauthoryear{{Vanderburg} et~al.,}{{Vanderburg}
  et~al.}{2015}]{Vanderburg2015}
{Vanderburg} A.,  et~al., 2015, \mn@doi [\nat] {10.1038/nature15527}, \href
  {http://adsabs.harvard.edu/abs/2015Natur.526..546V} {526, 546}

\bibitem[\protect\citeauthoryear{{Vauclair}, {Vauclair}  \&
  {Greenstein}}{{Vauclair} et~al.}{1979}]{Vauclair1979}
{Vauclair} G.,  {Vauclair} S.,   {Greenstein} J.~L.,  1979, \aap, \href
  {http://adsabs.harvard.edu/abs/1979A%26A....80...79V} {80, 79}

\bibitem[\protect\citeauthoryear{{Vennes}, {Kawka}  \& {N{\'e}meth}}{{Vennes}
  et~al.}{2010}]{Vennes2010}
{Vennes} S.,  {Kawka} A.,   {N{\'e}meth} P.,  2010, \mn@doi [\mnras]
  {10.1111/j.1745-3933.2010.00830.x}, \href
  {http://adsabs.harvard.edu/abs/2010MNRAS.404L..40V} {404, L40}

\bibitem[\protect\citeauthoryear{{Veras}}{{Veras}}{2016a}]{Veras2016}
{Veras} D.,  2016a, \mn@doi [Royal Society Open Science] {10.1098/rsos.150571},
  \href {https://ui.adsabs.harvard.edu/abs/2016RSOS....350571V} {3, 150571}

\bibitem[\protect\citeauthoryear{{Veras}}{{Veras}}{2016b}]{Veras2016b}
{Veras} D.,  2016b, \mn@doi [Monthly Notices of the Royal Astronomical Society]
  {10.1093/mnras/stw2170}, \href
  {https://ui.adsabs.harvard.edu/abs/2016MNRAS.463.2958V} {463, 2958}

\bibitem[\protect\citeauthoryear{{Veras} \& {Heng}}{{Veras} \&
  {Heng}}{2020}]{Veras2020b}
{Veras} D.,  {Heng} K.,  2020, \mn@doi [\mnras] {10.1093/mnras/staa1632}, \href
  {https://ui.adsabs.harvard.edu/abs/2020MNRAS.496.2292V} {496, 2292}

\bibitem[\protect\citeauthoryear{{Veras} \& {Scheeres}}{{Veras} \&
  {Scheeres}}{2020}]{VerasScheeres2020}
{Veras} D.,  {Scheeres} D.~J.,  2020, \mn@doi [\mnras] {10.1093/mnras/stz3565},
  \href {https://ui.adsabs.harvard.edu/abs/2020MNRAS.492.2437V} {492, 2437}

\bibitem[\protect\citeauthoryear{{Veras} \& {Wolszczan}}{{Veras} \&
  {Wolszczan}}{2019}]{Veras2019}
{Veras} D.,  {Wolszczan} A.,  2019, \mn@doi [\mnras] {10.1093/mnras/stz1721},
  \href {https://ui.adsabs.harvard.edu/abs/2019MNRAS.488..153V} {488, 153}

\bibitem[\protect\citeauthoryear{{Veras}, {Wyatt}, {Mustill}, {Bonsor}  \&
  {Eldridge}}{{Veras} et~al.}{2011}]{Veras2011}
{Veras} D.,  {Wyatt} M.~C.,  {Mustill} A.~J.,  {Bonsor} A.,   {Eldridge} J.~J.,
   2011, \mn@doi [\mnras] {10.1111/j.1365-2966.2011.19393.x}, \href
  {http://adsabs.harvard.edu/abs/2011MNRAS.417.2104V} {417, 2104}

\bibitem[\protect\citeauthoryear{{Veras}, {Mustill}, {Bonsor}  \&
  {Wyatt}}{{Veras} et~al.}{2013a}]{Veras2013a}
{Veras} D.,  {Mustill} A.~J.,  {Bonsor} A.,   {Wyatt} M.~C.,  2013a, \mn@doi
  [\mnras] {10.1093/mnras/stt289}, \href
  {http://adsabs.harvard.edu/abs/2013MNRAS.431.1686V} {431, 1686}

\bibitem[\protect\citeauthoryear{{Veras}, {Hadjidemetriou}  \& {Tout}}{{Veras}
  et~al.}{2013b}]{Veras2013b}
{Veras} D.,  {Hadjidemetriou} J.~D.,   {Tout} C.~A.,  2013b, \mn@doi [\mnras]
  {10.1093/mnras/stt1451}, \href
  {http://adsabs.harvard.edu/abs/2013MNRAS.435.2416V} {435, 2416}

\bibitem[\protect\citeauthoryear{{Veras}, {Jacobson}  \&
  {G{\"a}nsicke}}{{Veras} et~al.}{2014a}]{VerasJacobson2014}
{Veras} D.,  {Jacobson} S.~A.,   {G{\"a}nsicke} B.~T.,  2014a, \mn@doi [Monthly
  Notices of the Royal Astronomical Society] {10.1093/mnras/stu1926}, \href
  {https://ui.adsabs.harvard.edu/abs/2014MNRAS.445.2794V} {445, 2794}

\bibitem[\protect\citeauthoryear{{Veras}, {Shannon}  \& {G{\"a}nsicke}}{{Veras}
  et~al.}{2014b}]{VerasShannon2014}
{Veras} D.,  {Shannon} A.,   {G{\"a}nsicke} B.~T.,  2014b, \mn@doi [Monthly
  Notices of the Royal Astronomical Society] {10.1093/mnras/stu2026}, \href
  {https://ui.adsabs.harvard.edu/abs/2014MNRAS.445.4175V} {445, 4175}

\bibitem[\protect\citeauthoryear{{Veras}, {Carter}, {Leinhardt}  \&
  {G{\"a}nsicke}}{{Veras} et~al.}{2017}]{Veras2017}
{Veras} D.,  {Carter} P.~J.,  {Leinhardt} Z.~M.,   {G{\"a}nsicke} B.~T.,  2017,
  \mn@doi [\mnras] {10.1093/mnras/stw2748}, \href
  {https://ui.adsabs.harvard.edu/abs/2017MNRAS.465.1008V} {465, 1008}

\bibitem[\protect\citeauthoryear{{Veras}, {Xu}  \& {Rebassa-Mansergas}}{{Veras}
  et~al.}{2018}]{VerasXu2018}
{Veras} D.,  {Xu} S.,   {Rebassa-Mansergas} A.,  2018, \mn@doi [\mnras]
  {10.1093/mnras/stx2141}, \href
  {https://ui.adsabs.harvard.edu/abs/2018MNRAS.473.2871V} {473, 2871}

\bibitem[\protect\citeauthoryear{{Veras}, {Higuchi}  \& {Ida}}{{Veras}
  et~al.}{2019}]{VerasHiguchi2019}
{Veras} D.,  {Higuchi} A.,   {Ida} S.,  2019, \mn@doi [\mnras]
  {10.1093/mnras/stz421}, \href
  {https://ui.adsabs.harvard.edu/abs/2019MNRAS.485..708V} {485, 708}

\bibitem[\protect\citeauthoryear{{Veras}, {McDonald}  \& {Makarov}}{{Veras}
  et~al.}{2020}]{Veras2020a}
{Veras} D.,  {McDonald} C.~H.,   {Makarov} V.~V.,  2020, \mn@doi [\mnras]
  {10.1093/mnras/staa243}, \href
  {https://ui.adsabs.harvard.edu/abs/2020MNRAS.492.5291V} {492, 5291}

\bibitem[\protect\citeauthoryear{{Villaver} \& {Livio}}{{Villaver} \&
  {Livio}}{2007}]{Villaver2007}
{Villaver} E.,  {Livio} M.,  2007, \mn@doi [\apj] {10.1086/516746}, \href
  {http://adsabs.harvard.edu/abs/2007ApJ...661.1192V} {661, 1192}

\bibitem[\protect\citeauthoryear{{Villaver} \& {Livio}}{{Villaver} \&
  {Livio}}{2009}]{Villaver2009}
{Villaver} E.,  {Livio} M.,  2009, \mn@doi [\apjl]
  {10.1088/0004-637X/705/1/L81}, \href
  {http://adsabs.harvard.edu/abs/2009ApJ...705L..81V} {705, L81}

\bibitem[\protect\citeauthoryear{{Villaver}, {Livio}, {Mustill}  \&
  {Siess}}{{Villaver} et~al.}{2014}]{Villaver2014}
{Villaver} E.,  {Livio} M.,  {Mustill} A.~J.,   {Siess} L.,  2014, \mn@doi
  [\apj] {10.1088/0004-637X/794/1/3}, \href
  {http://adsabs.harvard.edu/abs/2014ApJ...794....3V} {794, 3}

\bibitem[\protect\citeauthoryear{{Voyatzis}, {Hadjidemetriou}, {Veras}  \&
  {Varvoglis}}{{Voyatzis} et~al.}{2013}]{Voyatzis2013}
{Voyatzis} G.,  {Hadjidemetriou} J.~D.,  {Veras} D.,   {Varvoglis} H.,  2013,
  \mn@doi [\mnras] {10.1093/mnras/stt137}, \href
  {https://ui.adsabs.harvard.edu/abs/2013MNRAS.430.3383V} {430, 3383}

\bibitem[\protect\citeauthoryear{{Wahhaj} et~al.,}{{Wahhaj}
  et~al.}{2013}]{Wahhaj2013}
{Wahhaj} Z.,  et~al., 2013, \mn@doi [\apj] {10.1088/0004-637X/773/2/179}, \href
  {https://ui.adsabs.harvard.edu/abs/2013ApJ...773..179W} {773, 179}

\bibitem[\protect\citeauthoryear{{Weiss} et~al.,}{{Weiss}
  et~al.}{2013}]{Weiss2013}
{Weiss} L.~M.,  et~al., 2013, \mn@doi [\apj] {10.1088/0004-637X/768/1/14},
  \href {https://ui.adsabs.harvard.edu/abs/2013ApJ...768...14W} {768, 14}

\bibitem[\protect\citeauthoryear{{Weiss} et~al.,}{{Weiss}
  et~al.}{2017}]{Weiss2017}
{Weiss} L.~M.,  et~al., 2017, \mn@doi [\aj] {10.3847/1538-3881/aa6c29}, \href
  {https://ui.adsabs.harvard.edu/abs/2017AJ....153..265W} {153, 265}

\bibitem[\protect\citeauthoryear{{Wilson}, {Farihi}, {G{\"a}nsicke}  \&
  {Swan}}{{Wilson} et~al.}{2019}]{Wilson2019}
{Wilson} T.~G.,  {Farihi} J.,  {G{\"a}nsicke} B.~T.,   {Swan} A.,  2019,
  \mn@doi [\mnras] {10.1093/mnras/stz1050}, \href
  {https://ui.adsabs.harvard.edu/abs/2019MNRAS.487..133W} {487, 133}

\bibitem[\protect\citeauthoryear{{Winn}, {Fabrycky}, {Albrecht}  \&
  {Johnson}}{{Winn} et~al.}{2010}]{Winn2010}
{Winn} J.~N.,  {Fabrycky} D.,  {Albrecht} S.,   {Johnson} J.~A.,  2010, \mn@doi
  [\apjl] {10.1088/2041-8205/718/2/L145}, \href
  {https://ui.adsabs.harvard.edu/abs/2010ApJ...718L.145W} {718, L145}

\bibitem[\protect\citeauthoryear{{Wittenmyer} et~al.,}{{Wittenmyer}
  et~al.}{2016}]{Wittenmyer2016}
{Wittenmyer} R.~A.,  et~al., 2016, \mn@doi [\apj] {10.3847/0004-637X/819/1/28},
  \href {https://ui.adsabs.harvard.edu/abs/2016ApJ...819...28W} {819, 28}

\bibitem[\protect\citeauthoryear{{Wright}, {Upadhyay}, {Marcy}, {Fischer},
  {Ford}  \& {Johnson}}{{Wright} et~al.}{2009}]{Wright2009}
{Wright} J.~T.,  {Upadhyay} S.,  {Marcy} G.~W.,  {Fischer} D.~A.,  {Ford}
  E.~B.,   {Johnson} J.~A.,  2009, \mn@doi [\apj]
  {10.1088/0004-637X/693/2/1084}, \href
  {https://ui.adsabs.harvard.edu/abs/2009ApJ...693.1084W} {693, 1084}

\bibitem[\protect\citeauthoryear{{Wyatt}, {Farihi}, {Pringle}  \&
  {Bonsor}}{{Wyatt} et~al.}{2014}]{Wyatt2014}
{Wyatt} M.~C.,  {Farihi} J.,  {Pringle} J.~E.,   {Bonsor} A.,  2014, \mn@doi
  [\mnras] {10.1093/mnras/stu183}, \href
  {http://adsabs.harvard.edu/abs/2014MNRAS.439.3371W} {439, 3371}

\bibitem[\protect\citeauthoryear{{Xu}, {Jura}, {Dufour}  \& {Zuckerman}}{{Xu}
  et~al.}{2016}]{Xu2016}
{Xu} S.,  {Jura} M.,  {Dufour} P.,   {Zuckerman} B.,  2016, \mn@doi [\apjl]
  {10.3847/2041-8205/816/2/L22}, \href
  {http://adsabs.harvard.edu/abs/2016ApJ...816L..22X} {816, L22}

\bibitem[\protect\citeauthoryear{{Xu} et~al.,}{{Xu} et~al.}{2018a}]{xu2018a}
{Xu} S.,  et~al., 2018a, \mn@doi [\mnras] {10.1093/mnras/stx3023}, \href
  {http://adsabs.harvard.edu/abs/2018MNRAS.474.4795X} {474, 4795}

\bibitem[\protect\citeauthoryear{{Xu} et~al.,}{{Xu} et~al.}{2018b}]{xu2018b}
{Xu} S.,  et~al., 2018b, \mn@doi [\apj] {10.3847/1538-4357/aadcfe}, \href
  {https://ui.adsabs.harvard.edu/abs/2018ApJ...866..108X} {866, 108}

\bibitem[\protect\citeauthoryear{{Zhou} et~al.,}{{Zhou}
  et~al.}{2015}]{Zhou2015}
{Zhou} G.,  et~al., 2015, \mn@doi [\apjl] {10.1088/2041-8205/814/1/L16}, \href
  {https://ui.adsabs.harvard.edu/abs/2015ApJ...814L..16Z} {814, L16}

\bibitem[\protect\citeauthoryear{{Zuckerman}, {Koester}, {Reid}  \&
  {H{\"u}nsch}}{{Zuckerman} et~al.}{2003}]{Zuckerman2003}
{Zuckerman} B.,  {Koester} D.,  {Reid} I.~N.,   {H{\"u}nsch} M.,  2003, \mn@doi
  [\apj] {10.1086/377492}, \href
  {http://adsabs.harvard.edu/abs/2003ApJ...596..477Z} {596, 477}

\bibitem[\protect\citeauthoryear{{Zuckerman}, {Koester}, {Melis}, {Hansen}  \&
  {Jura}}{{Zuckerman} et~al.}{2007}]{Zuckerman2007}
{Zuckerman} B.,  {Koester} D.,  {Melis} C.,  {Hansen} B.~M.,   {Jura} M.,
  2007, \mn@doi [\apj] {10.1086/522223}, \href
  {https://ui.adsabs.harvard.edu/abs/2007ApJ...671..872Z} {671, 872}

\bibitem[\protect\citeauthoryear{{Zuckerman}, {Melis}, {Klein}, {Koester}  \&
  {Jura}}{{Zuckerman} et~al.}{2010}]{Zuckerman2010}
{Zuckerman} B.,  {Melis} C.,  {Klein} B.,  {Koester} D.,   {Jura} M.,  2010,
  \mn@doi [\apj] {10.1088/0004-637X/722/1/725}, \href
  {http://adsabs.harvard.edu/abs/2010ApJ...722..725Z} {722, 725}

\bibitem[\protect\citeauthoryear{{von Hippel}, {Kuchner}, {Kilic}, {Mullally}
  \& {Reach}}{{von Hippel} et~al.}{2007}]{vonHippel2007}
{von Hippel} T.,  {Kuchner} M.~J.,  {Kilic} M.,  {Mullally} F.,   {Reach}
  W.~T.,  2007, \mn@doi [\apj] {10.1086/518108}, \href
  {http://adsabs.harvard.edu/abs/2007ApJ...662..544V} {662, 544}

\makeatother
\end{thebibliography}

\bsp	
\label{lastpage}
\end{document}